\documentclass{aa}  

\usepackage{graphicx}	
\usepackage{amsmath}	
\usepackage{lipsum}   
\usepackage{longtable}
\usepackage{subcaption}
\usepackage{array, multirow}
\usepackage{xcolor}
\usepackage{comment}  
\usepackage{threeparttablex}
\usepackage{booktabs}

\usepackage{xurl}

\usepackage{txfonts}
\usepackage{xspace}
\newcommand{\source}{MAXI J1834--021\xspace}
\newcommand{\unitL}{erg~s$^{-1}$\xspace}
\newcommand{\unitF}{erg~cm$^{-2}$~s$^{-1}$\xspace}
\newcommand\T{\rule{0pt}{2.6ex}}       
\newcommand\B{\rule[-1.2ex]{0pt}{0pt}}

\usepackage{natbib,twoopt}
\usepackage[breaklinks, colorlinks]{hyperref} 
\usepackage[hyphenbreaks]{breakurl}
\DeclareMathAlphabet\mathzapf       {T1}{pzc} {mb} {it}
\bibpunct{(}{)}{;}{a}{}{,}             
\definecolor{cobalt}{rgb}{0.06, 0.2, 0.65}
\hypersetup{
  colorlinks,
  citecolor=cobalt,
  linkcolor=cobalt,
  urlcolor=cobalt,
}
\makeatletter
  \newcommandtwoopt{\citeads}[3][][]{\href{http://adsabs.harvard.edu/abs/#3}%
    {\def\hyper@linkstart##1##2{}%
     \let\hyper@linkend\@empty\citealp[#1][#2]{#3}}}
  \newcommandtwoopt{\citepads}[3][][]{\href{http://adsabs.harvard.edu/abs/#3}%
    {\def\hyper@linkstart##1##2{}%
     \let\hyper@linkend\@empty\citep[#1][#2]{#3}}}
  \newcommandtwoopt{\citetads}[3][][]{\href{http://adsabs.harvard.edu/abs/#3}%
    {\def\hyper@linkstart##1##2{}%
     \let\hyper@linkend\@empty\citet[#1][#2]{#3}}}
  \newcommandtwoopt{\citeyearads}[3][][]%
    {\href{http://adsabs.harvard.edu/abs/#3}
    {\def\hyper@linkstart##1##2{}%
     \let\hyper@linkend\@empty\citeyear[#1][#2]{#3}}}
\makeatother

\def\arc{\mbox{$^{\prime\prime}$}}

\def\deg{\mbox{$^{\circ}$}\xspace}

\def\nustar {\emph{NuSTAR}\xspace}
\def\swift {\emph{Swift}\xspace}
\def\nicer {\emph{NICER}\xspace}
\def\XMM {\emph{XMM-Newton}\xspace}

\begin{document}

   \title{On the nature of the X-ray binary transient \source: clues from its first observed outburst}


 \author{
A.~Manca\inst{1}
\and 
A.~Marino\inst{2,3,4}
\and 
A.~Borghese\inst{5,6,7,8}
\and 
F.~Coti Zelati\inst{2,3,9}
\and
G.~Mastroserio\inst{10}
\and
A.~Sanna\inst{1}
\and
J.~Homan\inst{11}
\and
R.~Connors\inst{12}
\and
M.~Del Santo\inst{4}
\and
M.~Armas Padilla\inst{5,6}
\and
T.~Mu\~{n}oz-Darias\inst{5,6}
\and
T.~Di~Salvo\inst{13}
\and
N.~Rea\inst{2,3}
\and
J.~A.~Garc\'{i}a\inst{14,15}
\and
A.~Riggio\inst{1,16}
\and
M.~C. Baglio\inst{9}
\and
L.~Burderi\inst{1,4}
}

\institute{
Universit\`a degli Studi di Cagliari, Dipartimento di Fisica, SP Monserrato-Sestu km 0.7, I-09042 Monserrato, Italy
\email{arianna.manca@inaf.it}
\and
Institute of Space Sciences (ICE, CSIC), Campus UAB, Carrer de Can Magrans s/n, E-08193 Barcelona, Spain
\and
Institut d'Estudis Espacials de Catalunya (IEEC), E-08034 Barcelona, Spain
\and
INAF/IASF Palermo, via Ugo La Malfa 153, I-90146 Palermo, Italy
\and
Instituto de Astrofísica de Canarias (IAC), Vía Láctea s/n, La Laguna 38205, S/C de Tenerife, Spain
\and
Departamento de Astrofísica, Universidad de La Laguna, La Laguna, E-38205, S/C de Tenerife, Spain
\and
INAF-Osservatorio Astronomico di Roma, Via Frascati 33, I-00040 Monte Porzio Catone (RM), Italy
\and
European Space Agency (ESA), European Space Astronomy Center (ESAC), Camino Bajo del Castillo s/n, 28692 Villafranca del Castillo, Madrid, Spain
\and
INAF, Osservatorio Astronomico di Brera, via E. Bianchi 46, I-23807 Merate (LC), Italy
\and
Dipartimento di Fisica, Universit\`a Degli Studi di Milano, Via Celoria 16, 20133 Milano, Italy
\and
Eureka Scientific, Inc., 2452 Delmer Street, Oakland, CA 94602, USA 
\and
Department of Physics, Villanova University, 800 Lancaster Avenue, Villanova, PA 19085, USA
\and
Universit\`a degli Studi di Palermo, Dipartimento di Fisica e Chimica - Emilio Segr\`e, via Archirafi 36, I-90123 Palermo, Italy
\and
X-ray Astrophysics Laboratory, NASA Goddard Space Flight Center, Greenbelt, MD 20771, USA
\and
Cahill Center for Astronomy and Astrophysics, California Institute of Technology, Pasadena, CA 91125, USA
\and
INFN, Sezione di Cagliari, Cittadella Universitaria, 09042 Monserrato, CA, Italy
}

   \date{Received XXX; accepted XXX}

 
  \abstract
   {\source is a new X-ray transient that was discovered in February 2023. We analysed the spectral and timing properties of \source using \nicer, \nustar and \swift data collected between March and October 2023. The light curve showed a main peak followed by a second activity phase. The majority of the spectra extracted from the individual \nicer observations could be adequately fitted with a Comptonisation component alone, while a few of them required an additional thermal component. 
   The spectral evolution is consistent with a softening trend as the source gets brighter in X-rays. We also analysed the broadband spectrum combining data from simultaneous \nicer and \nustar observations on 2023 March 10. This spectrum can be fitted with a disc component with a temperature at the inner radius of $kT_{\rm in} \sim 0.4$~keV and a Comptonisation component with a power-law photon index of $\Gamma \sim 1.8$. 
   By including a reflection component in the modelling, we obtained a 3$\sigma$ upper limit for the inner disc radius of 11.4 gravitational radii. 
We also detected a quasi-periodic oscillation (QPO), whose central frequency varies with time (from 2\,Hz to $\sim$0.9\,Hz) and anti-correlates with the hardness ratio.
Based on the observed spectral-timing properties, \source, can be classified as a low-mass X-ray binary in outburst. However, we are not able to draw a definitive conclusion on the nature of the accreting compact object, which at the moment could as well be a black hole or a neutron star.}



   \maketitle
%

\section{Introduction}

X-ray binary systems are roughly classified into two major sub-classes: Low-Mass X-ray Binaries \citep[LMXBs;][]{TOO2010, Bahramian2023, DiSalvo2023} if the mass of the companion star is $\lesssim 1 M_\odot$ \citep{Lewin1980}, and High-Mass X-ray Binaries (HMXBs) if the companion star mass is $\geq 8 M_\odot$ \citep{Fortin2023}. The two classes differ also in their accretion mechanisms: in LMXBs, accretion occurs through Roche-lobe overflow, when the companion star fills its Roche lobe and the matter is pulled towards the compact object, with the formation of an accretion disc. In HMXBs, accretion mainly takes place directly through the stellar wind of the companion star \citep[see][and references therein]{Karino2019}. The compact object can be either a black hole (BH) or a neutron star (NS). The presence of a NS can be inferred by the detection of peculiar features arising from the existence of a solid surface, such as type-I X-ray bursts---uncontrolled burning of fuel on the NS surface \citep[see, e.g.,][for a review]{Galloway2021}---or X-ray pulsations \citep[see, e.g.,][]{DiSalvo2022}. However, the absence of these features does not necessarily imply the presence of a BH.


LMXBs can be classified into persistent and transient systems. Persistent systems are characterised by a steady accretion rate and emit stable X-ray emission, sometimes with slight variations in intensity. Transient systems, on the other hand, go through cycles of low activity (quiescence) and high activity (outbursts). During quiescence, the X-ray luminosity drops below $10^{33}$~\unitL. However, during sudden outbursts, the luminosity can reach values of $10^{36-38}$~\unitL (see \citealt[for reviews on BH LMXBs]{Belloni2016, Motta2021}, and see \citealt[for a review on NS LMXBs]{DiSalvo2023}). The difference between persistent and transient sources could be due to either the timescale of these transitions---which might be too long to observe in persistent sources---or the mass accretion rate, which remains stable enough in persistent systems to support continuous X-ray emission, while it changes significantly in transient systems, leading to intermittent activity \citep[see, e.g.,][and references therein for examples]{Degenaar2014, Degenaar2017}.

The spectral and timing properties of LMXBs 
enable the identification of various accretion states. They are studied through energy spectra, Power Density Spectra (PDS) and Hardness/Intensity Diagrams (HIDs). The energy spectra show two main components: a thermal multi-colour disc, 
and a Comptonised component that can usually be modelled by a power-law with a high-energy cutoff. In the case of NS LMXBs, an additional black-body-like component may emerge, caused by thermal emission from the NS solid surface or a boundary layer between the disc and the NS. The spectra of both BH and NS binaries can also show an Fe K$\alpha$ emission line at $\sim 6.4-6.7$~keV, sometimes relativistically broadened, due to fluorescence or recombination processes in the inner disc. A reflection hump can appear at energies $\sim 20$~keV \citep[see][for more details and examples of NS systems]{Fabian1989, Bhattacharyya2007, Cackett2008}. 

The states of BH LMXBs can be roughly identified according to their position in the HID. The outburst begins with the source leaving the quiescence state and entering a low luminosity/hard state (LHS). This state is characterised by a predominance of the Comptonisation component. As the spectrum softens, the source enters a hard/intermediate state (HIMS), where the thermal component begins to arise and the hard power-law component becomes steeper. The HIMS is followed by the soft/intermediate state (SIMS) moving farther to the left of the HID, in which the spectrum becomes slightly softer. In the high luminosity/soft state (HSS), the spectrum becomes dominated by the thermal accretion disc component, which peaks at temperatures of $\sim 1$~keV.  By the end of the outburst, the source enters again the HIMS at lower luminosities than during the hard-to-soft transition, and eventually enters the LHS again and reaches quiescence. This hysteresis cycle was first reported by \citet{Miyamoto1995} for BH candidate LMXBs. 
The relationship between X-ray luminosity and state transition, though, is not straightforward, as different systems reach equal X-ray luminosities in different spectral states and, therefore, the X-ray luminosity at the transition is not the same for all sources. Some sources do not complete the transition to soft states. These \emph{failed-transition} outbursts are divided into two groups: the outbursts in which the source transitions to the HIMS, but never reaches the HSS; and the ones in which the source does not transition to the HIMS, but peaks in the LHS \citep{Alabarta2021}. Failed-transition outbursts have been observed in 
both BH \citep{Capitanio2009, DelSanto2016, Bassi2019, Garcia2019, Wang2022} and NS \citep[see, e.g.,][]{Marino2022, Manca2023a, Manca2023b} systems. In transient BH binaries, these outbursts are often referred to as `hard-only' or `failed-transition' outburst and they have been observed in about 40\% of the sources in this class \citep{Alabarta2021}. These outbursts are generally fainter than those featuring complete hard-to-soft transitions \citep{Tetarenko2016}. Although still not fully understood, the `failed state transition' behaviour is thought to result from the mass accretion rate not reaching a specific threshold \citep[see, e.g.,][]{Esin1997, Marcel2022}, and suggested to be about 0.1 $L_{\rm Edd}$ \citep{Tetarenko2016}, where $L_{\rm Edd}$ is the Eddington luminosity for a stellar-mass BH, or alternatively from the mass distribution in the accretion disc \citep{Campana2013}. 
Quasi-periodic oscillations (QPOs) can appear in the PDS in the different spectral states as narrow, discrete features. They have been hypothesised to be related to characteristic periodicities of the accretion flow, such as the keplerian motion of matter in the disc and the precession motions \citep[see, e.g.,][for one possible interpretation]{Ingram2009, IngramMotta2014}. They are usually divided into high-frequency QPOs and low-frequency QPOs, the latter type in turn divided into Type-A, B, and C (\citealt{Wijnands1999, Remillard2002}; see \citealt{IngramMotta2019} for a review). Type-C QPOs are seen in the HIMS and LHS, while Type-B and A are only observed during the transition to the soft state. 


NS LMXBs exhibit spectral properties similar to those of BHs, but with some differences due to the presence of a surface. NS LMXBs are generally classified into two categories: Z and Atoll sources. This classification is based on the track they follow in a Colour-Colour Diagram (CCD) during an outburst, as well as their timing features \citep{Vanderklis1989}. Z sources typically have high mass accretion rates and correspondingly high luminosities $\geq 0.5 L_{\rm Edd}$. Atoll sources, on the other hand, have lower luminosities $0.01-0.5 L_{\rm Edd}$ \citep{Lewin1990}. Therefore, this classification likely arises from differences in accretion rates, even though the underlying processes are similar \citep[see e.g.][]{Homan2010}.
For Z sources, the CCD is divided into three branches: the horizontal, normal and flaring branch. During the activity phase, the source moves continuously along the Z-shaped track in the CCD. Atoll sources states are classified into island, lower, and upper banana states. The outburst in Atoll sources evolves from the hard island states to the banana branch \citep[see, e.g.,][]{Altamirano2008}. NS systems have a more diverse QPO phenomenology and their classification is more complex.
In Z sources, low-frequency QPOs have been classified into three main types: horizontal, normal and flaring branch oscillations. They are believed to correspond to Type-C, B and A QPOs of BHs, respectively \citep{casella2005}. The same classification can be applied to Atoll systems, which show only horizontal and flaring branch oscillation-like  QPOs \citep{motta2017}. Low-frequency QPOs may appear in low and intermediate states \citep[see][for a full review]{VanderKlis2006}.
A hysteresis-like spectral evolution is also observed in NS LMXBs in persistent and transient Atoll sources \citep[see][]{MunozDarias2014}. The duration of this cycle generally matches the length of the outburst ($\sim 1$~month--year). However, in persistent sources or transients with long outbursts, multiple cycles can occur in succession. Similar to BH LMXBs, in the HID, this cycle progresses in an anticlockwise direction, with the transition from hard to soft states occurring at the point of highest luminosity.

\source is an X-ray binary whose exact classification is still subject of debate. It was detected for the first time on 2023 February 5 (MJD 59980) by MAXI/GSC \citep{Atel15929} as a new X-ray transient. The 2--10~keV X-ray flux was observed to increase during the outburst for a period of 10--20 days before decreasing again \citep{Atel15946}. 
No radio counterpart was found in AMI-LA observations performed at a central frequency of 15.5~GHz \citep{Atel15939}.


In this work, we analyse the 2023 outburst of \source with \nicer, \nustar, and \swift data. We outline the spectral properties of the outburst by individually analysing the \nicer and \swift observations, and analyse a broadband \nicer and \nustar spectrum in order to investigate the presence of a reflection component.
Moreover, we conduct a timing analysis of the \nicer and \nustar observations, to determine the temporal variability and investigate a possible QPO component.

\section{Observations and data reduction}

\source was monitored from March 6 to October 2, 2023, with the Neutron Star Interior Composition Explorer (\nicer; \citealt{NICER2016}) for a total exposure of 142 ks split in 81 observations, and with \swift-XRT for a total exposure time of $\sim$73\,ks split in 42 observations. Moreover, the Nuclear Spectroscopic Telescope Array (\nustar; \citealt{harrison13}) observed the source on March 10, 2023 for an elapsed time of $\sim$55\,ks and an on-source exposure time of $\sim$29\,ks. Table \ref{tab:new_obs} reports the list of the observations analysed in this work, with their respective start and stop dates, exposures, and source count rates.

Data reduction was performed using tools incorporated in \textsc{heasoft 6.31}. In the following, all
uncertainties are quoted at 90\%  confidence level (c.l.) unless otherwise stated.




\subsection{\nicer}
\nicer data were processed with the \texttt{nicerl2} pipeline, with default screening settings: i) exclusion of time intervals in the proximity of the South Atlantic Anomaly; ii) elevation angle of at least 30$^\circ$ over the Earth's limb; iii) minimum angle of 40$^\circ$ from the bright Earth limb; iv) maximum angular distance between the source direction and \nicer pointing direction of 0.015$^\circ$. Upon visual inspection, several \nicer light curves showed irregular increases in count rates throughout the outburst. We ascertained that each of them corresponded to overshoots reaching values outside the recommended range, probably due to charged particle contamination\footnote{\url{https://heasarc.gsfc.nasa.gov/docs/nicer/analysis_threads/overshoot-intro/}}. We decided to cut the contaminated sections of the observations. We produced the spectra and background files with the \texttt{nicerl3-spect} pipeline, using the \textsc{scorpeon} default model for the creation of the background file. We rebinned the data with \textsc{ftgrouppha} following the Kaastra \& Bleeker optimal binning algorithm \citep{KaastraBleeker2016} in order to have at least 20 counts per bin. 

\subsection{\nustar}
\nustar data were processed and analysed using the NuSTAR Data Analysis Software (\texttt{nustardas}) and \texttt{caldb} v.20230404. The tool \texttt{nupipeline} was employed with default options to create cleaned event files and filter out time intervals associated with the satellite passing through the South Atlantic Anomaly. For both focal plane modules (FPMs), source photons were collected within a circle of radius 90 arcsec centred on the source position, while background photons were extracted from a circle of radius 180 arcsec far from the source. \source was detected up to an energy of $\approx$60\,keV in both FPMs. Background-subtracted spectra, instrumental responses, and auxiliary files were created using the tool \texttt{nuproducts}. Spectra of both FPMs were grouped to have at least 100 counts per energy channel.

\subsection{\swift}
The \swift single exposure ranged from 0.2 to 3.4\,ks, with the XRT operating either in windowed timing (WT; readout time of 1.77\,ms) or photon counting (PC; 2.51\,s) modes. We processed the data adopting standard cleaning criteria and created exposure maps with the task \texttt{xrtpipeline}. For the spectral analysis, we selected events with grades 0--12 and 0 for PC and WT data, respectively. We accumulated the source counts from a circular region with a radius of 20 pixels (1 pixel = 2.36 arcsec). To evaluate the background, we extracted the events within an annulus centred on the source position with radii of 40--80 pixels and 80--120 pixels for PC-mode and WT-mode observations, respectively. In case a pointing performed in PC mode was affected by pile-up, we followed the online analysis thread\footnote{\url{https://www.swift.ac.uk/analysis/xrt/pileup.php}} to determine the size of the core of the point-spread function to be excluded from our analysis. We generated the spectra with the corresponding ancillary response files through \texttt{xselect} and the \texttt{xrtmkarf} tool. The response matrices version 20131212v015 and 20130101v014 available in the XRT calibration database were assigned to each spectrum in WT and PC mode, respectively. The \swift-XRT background-subtracted spectra were grouped according to a variable minimum number of counts depending on the available counting statistics, varying between 10 and 40 counts per spectral bin.

\section{Data analysis and results}

\begin{figure*}
    \centering
    \includegraphics[width=\textwidth]{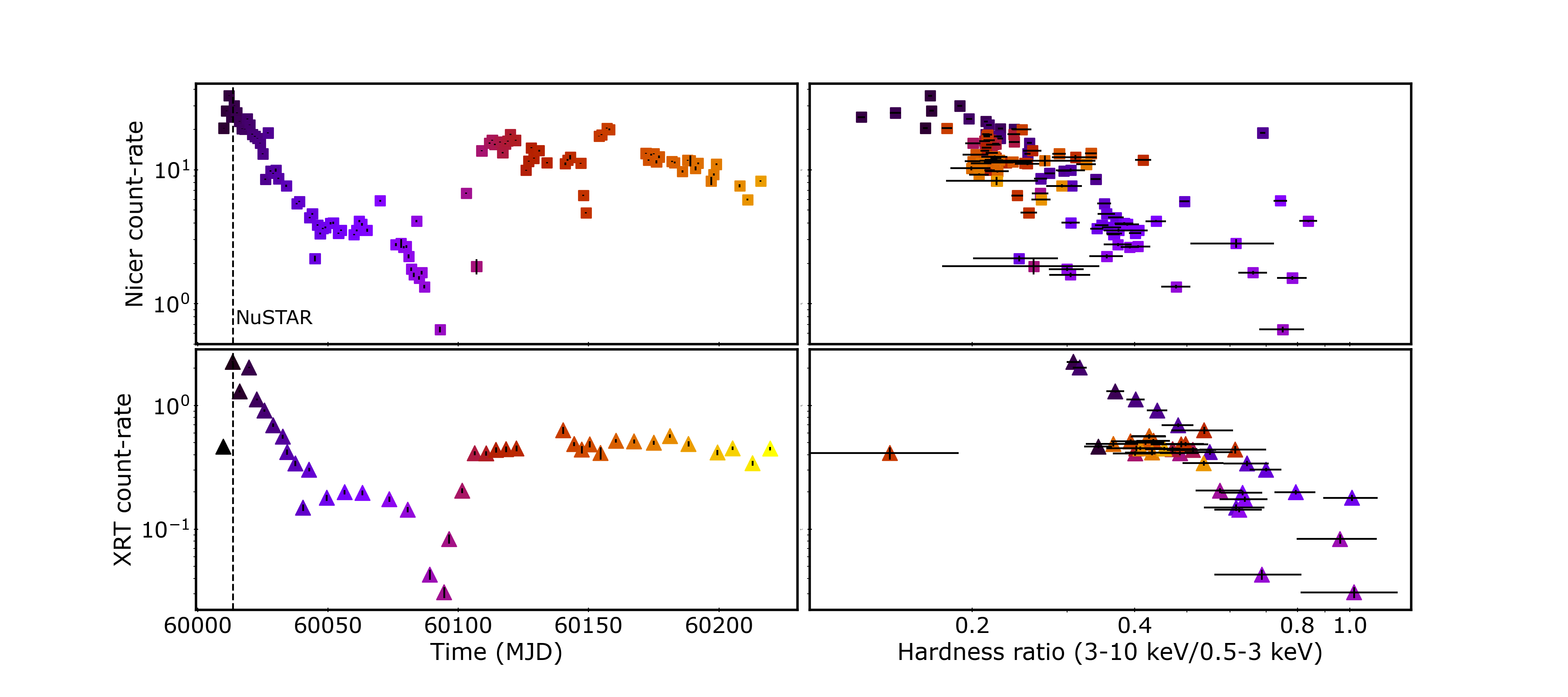}
    \caption{Light curve (left panel) and HID (right panel) extracted from \nicer (top) and \swift-XRT (bottom). The count-rates were calculated in the 0.5--10~keV energy range, while the hardness ratios were derived considering the 0.5--3 and 3--10~keV energy bands for both instruments. The colour gradient indicates the temporal sequence. The vertical dashed lines in the left panels denote the epoch of the \nustar\ observation.}
    \label{fig:HID}
\end{figure*}

\subsection{Source position}
\label{sec:position}

We calculated the XRT position of \source using the \swift UKSSDC’s online tool\footnote{\url{https://www.swift.ac.uk/user_objects/}} with the UVOT-enhanced method \citep{evans09}. We derived R.A. = 18$^\mathrm{h}$36$^\mathrm{m}$09$\fs$09, Decl. = -02$^{\circ}$17$^{\prime}$15$\farcs$6 (J2000.0), with an error radius of 2.4\arc\ at the 90\% confidence level.
This updated position does not match any optical source listed in Gaia DR3 \citep{GaiaDR32023}, nor the two optical sources detected in prompt observations using the Las Cumbres Observatory and Faulkes \citep{Atel15940} at a 90\% c.l.. The closest optical source, the faint one with a magnitude $G$ of 20.8, is 2.5 arcsec away and thus marginally consistent with this position.

\subsection{Outburst light curve and hardness intensity diagram}
\label{sec:HID}
Figure \ref{fig:HID}, on the left panel, shows the light curve of the outburst extracted using data from \nicer and \swift.
It indicates a morphological complex outburst, starting with the peak of the outburst and followed by an $\sim 100$ days long decrease, and then a rebrightening phase of similar length. The highest count rate recorded by \nicer is $\sim 30$~cts~s$^{-1}$ on MJD 60014, followed by a decrease to the level of $\sim 5$~cts~s$^{-1}$ and subsequent rise to $\sim 20$~cts~s$^{-1}$. The level remains almost steady between 10 and 20~cts~s$^{-1}$ for about 2--3 months, slowly decreasing to $\sim 10$~cts~s$^{-1}$ around October 2, 2023, end date of the observations.
We did not detect any type-I burst in our data. As shown in Figure\,\ref{fig:HID}, on the right panel, we also derive the hardness ratio, with the soft band defined as the range 0.5--3~keV, and the hard band as the range 3--10~keV. In the HID,  we can see that most of the observations stack at similar hardness ratio values, e.g., at $\sim$0.2--0.4 for \nicer, for varying count rates. Only observations taken when the source was at its lowest luminosity significantly deviate from the above trend and show an increase in hardness.

\subsection{Spectral analysis}
For all instruments, we performed the spectral analysis within \textsc{xspec} 12.13.1 \citep{Arnaud1996}. The interstellar absorption is modelled through the \texttt{TBabs} component, with the abundances provided by \citet{Wilms2000} and cross sections of \citet{Verner}. We derived the 
unabsorbed fluxes with the \texttt{cflux} convolution model.

\subsubsection{\nicer and \swift monitoring} 
\label{sec:spectral}

\source was monitored with an almost daily cadence with \nicer during 2023. In this paper, we report on the analysis of the first 81 pointings (see Table \ref{tab:new_obs}). We analysed each spectrum separately, keeping data in a range over which the source signal would lie above the background. This range varies between 0.5--8 keV and 0.5--5.5 keV, depending on the brightness of the source.
All spectra were analysed using three models: one including only the Comptonisation spectrum (Model 0, \texttt{TBabs$\times$nthComp}); the other two characterised also by the presence of an additional thermal component, which was in one case a black body (Model 1A, \texttt{TBabs$\times$(nthComp + bbodyrad)}), and in the other a disc black body (Model 1B, \texttt{TBabs$\times$(nthComp + diskbb)}). The first scenario would be appropriate for a NS LMXB where the thermal emission arises from either the NS surface or the boundary layer surrounding it, while the second scenario, where the thermal emission comes from the accretion disc, would apply to both BH and NS LMXBs. The outcome of an F-test on the statistical significance of adding the thermal component was used as a criterion to choose between Model 0 and Model 1A or 1B. In particular, whenever the probability of improvement by chance of the \texttt{bbodyrad} or \texttt{diskbb} component was estimated to be lower than 10$^{-4}$, we considered the thermal emission significant and chose Model 1A/1B instead of Model 0. When the thermal component was found to be statistically significant, we tied together the black body temperature $kT_{\rm bbody}$, or disc black body temperature $kT_{\rm disc}$, with the seed photons temperature $kT_{\rm seed}$, since the fits could not constrain them both when left free and independent. In all cases, we fixed the hydrogen column density $N_{\rm H}$ to 0.9$\times10^{22}$\,cm$^{-2}$ (i.e., the value obtained from the broadband spectral analysis adopting Model 3, see Sec.\,\ref{sec:broadband}). The \texttt{inp\_type} parameter was set to 0 for Model 1A, and 1 for Model 1B,  to distinguish between black-body-like and disc-black-body-like distributions of seed photons, respectively. Additionally, since the lack of hard X-rays coverage prevented us from constraining the high energy cutoff, the electron temperature $kT_{\rm e}$ was fixed to 100 keV in all models. 
Moreover, \source was intensively monitored by \swift-XRT as well. In order to increase the statistics, we merged observations carried out a few days apart (see Table\,\ref{tab:new_obs}), ending up with 40 spectra. We fit the spectra simultaneously with Model 0. All parameters were left free to vary between the datasets except $N_{\rm H}$, which
was frozen at 0.9$\times10^{22}$\,cm$^{-2}$. $kT_{\rm seed}$ was also not constrained by the fit, so that we fixed it at 0.1\,keV. 
The \swift observations did not require an additional thermal component, most likely because of the low photon statistics of the single \swift-XRT observations.

\begin{figure*}
\centering
\includegraphics[width=1.\columnwidth]{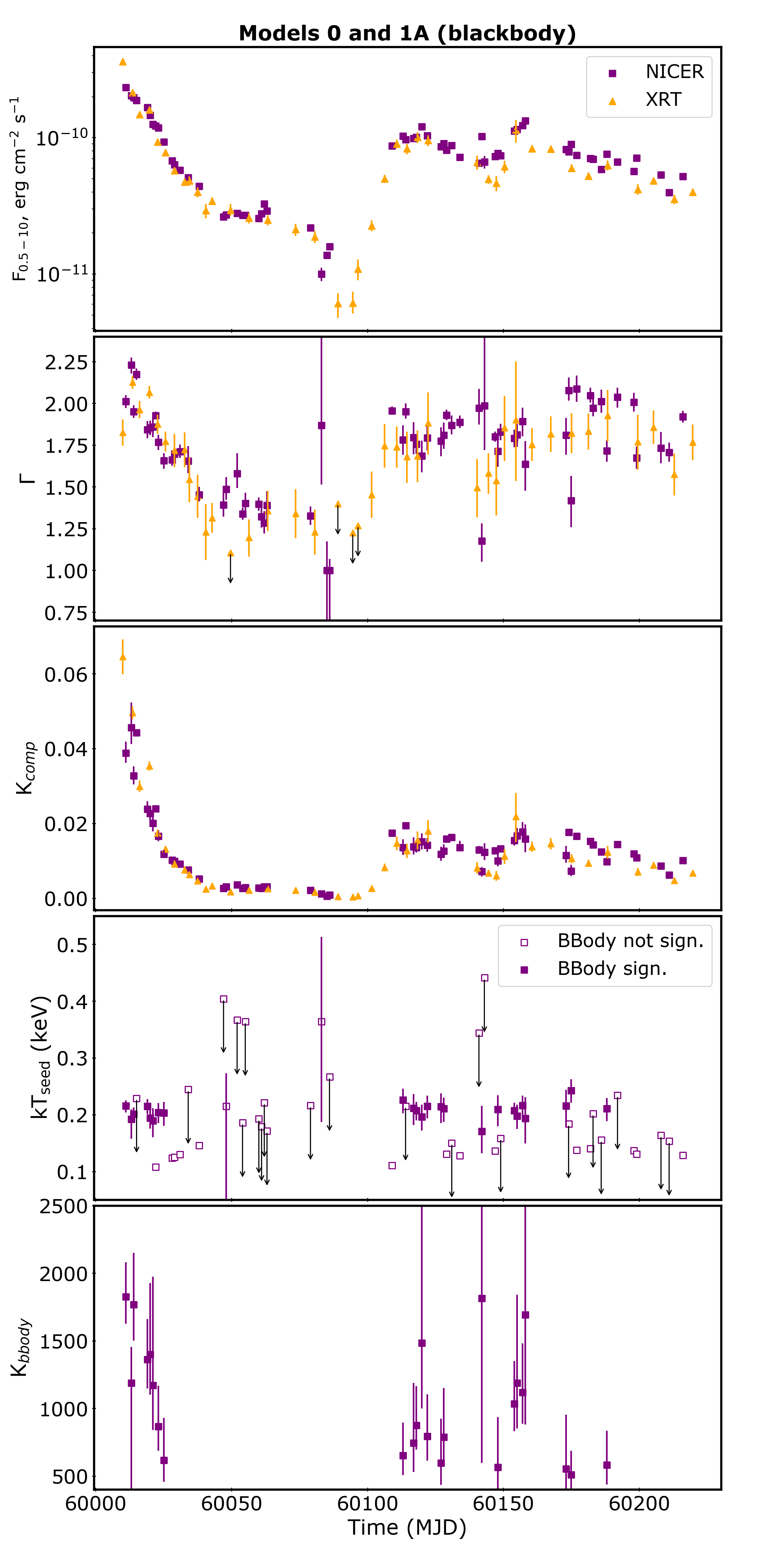} 
\includegraphics[width=1.\columnwidth]{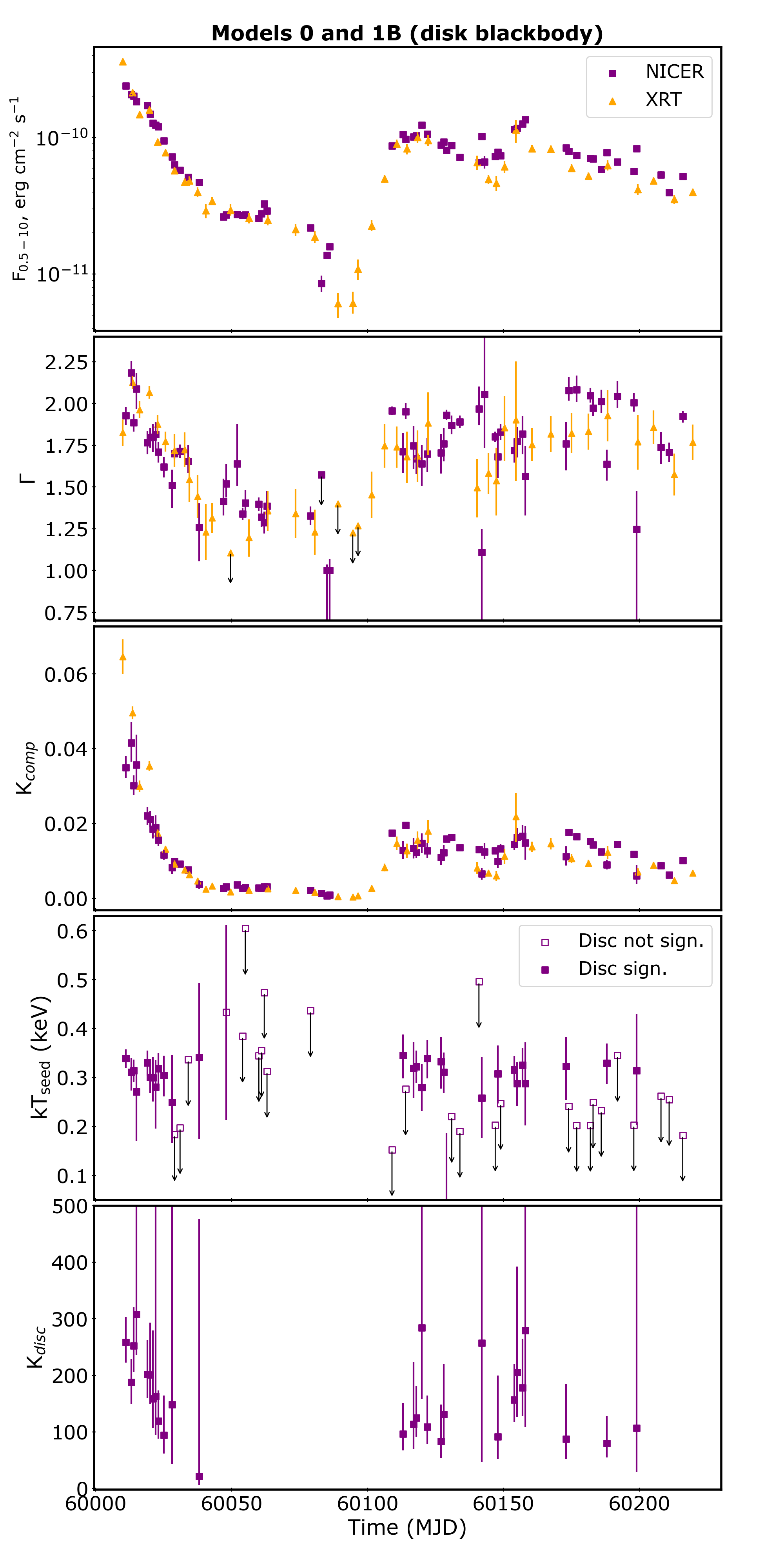} 
\vspace{-0.3cm}
\caption{\label{fig:tower-plot-nicer} Evolution of the main spectral parameters from the 2023 \nicer and \swift campaign on \source with Models 0 and 1A (left) and 1B (right). The top panels report the unabsorbed flux in the 0.5--10\,keV range as estimated from spectral fitting for both \nicer (purple squares) and \swift-XRT (orange triangles). The evolution of the main best-fit parameters used in the spectral analysis (Sec.\,\ref{sec:spectral}) is plotted in the other panels. In the fourth panels, we used filled and empty squares to differentiate between the observations where the inclusion of the \texttt{bbodyrad}/\texttt{diskbb} component is significant (in this case $kT_{\rm seed}=kT_{\rm bbody}$ or $kT_{\rm seed}=kT_{\rm disc}$) and observations where it is not significant.}
\end{figure*}

\begin{figure}
\centering
\includegraphics[width=1.\columnwidth]{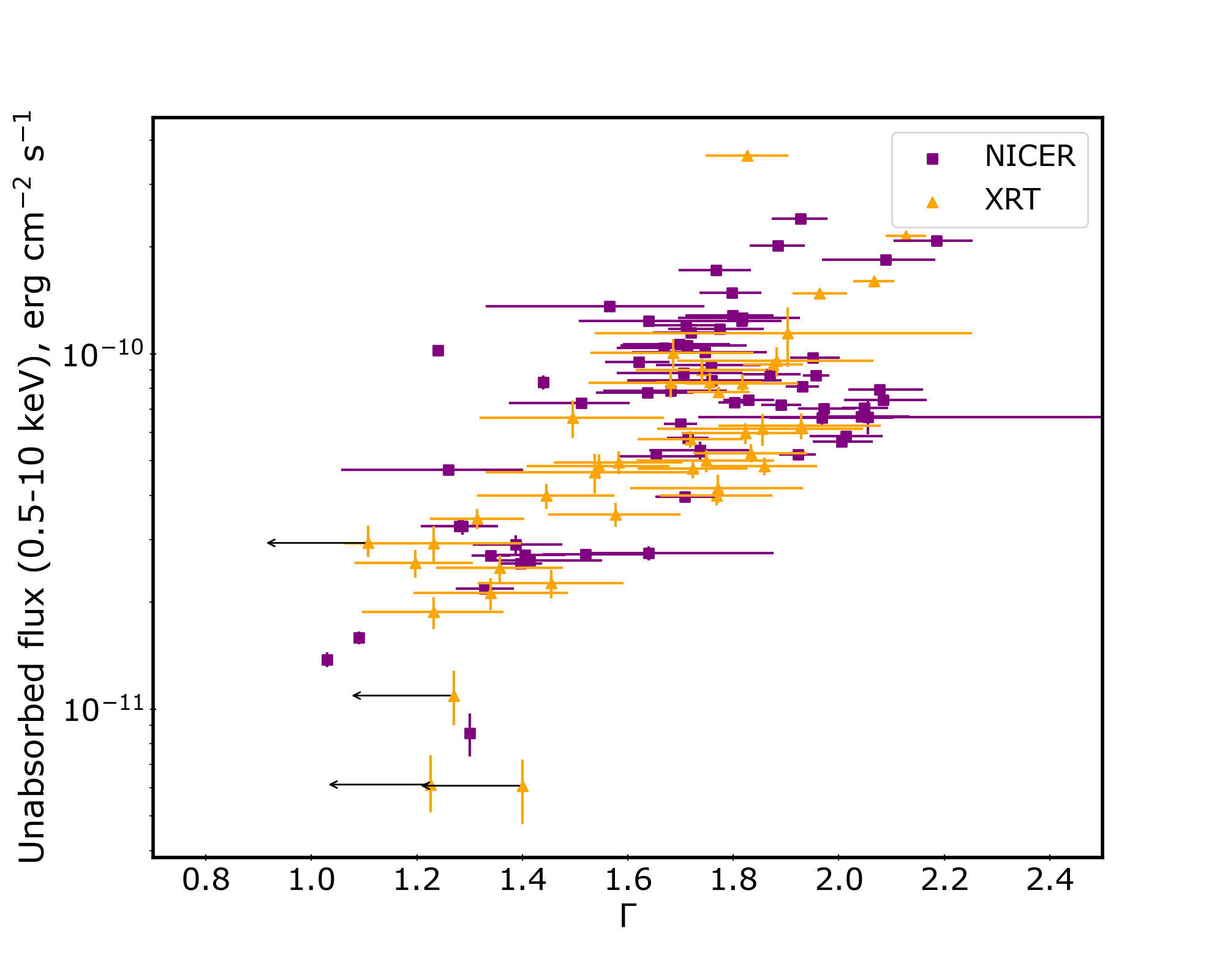} 
\vspace{-0.3cm}
\caption{\label{fig:flux-gamma} Evolution of the best-fit power-law photon index $\Gamma$ vs. unabsorbed 0.5--10\,keV flux for all \nicer (purple squares) and \swift-XRT (orange triangles) observations. The data correspond to the results of the fits using Models 0-1B.}
\end{figure}

\begin{figure*}
\centering
\includegraphics[width=0.95\columnwidth]{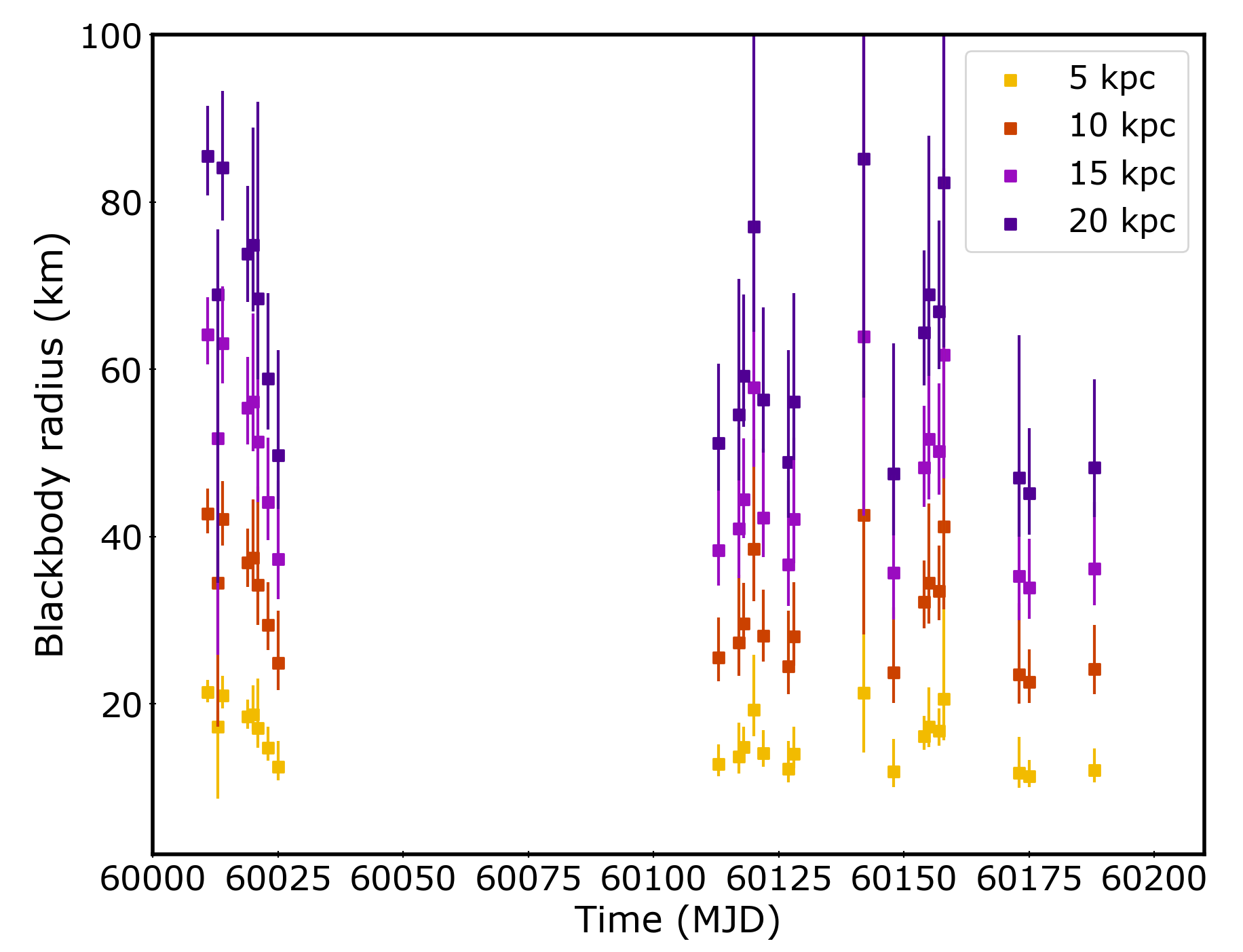} 
\includegraphics[width=1.\columnwidth]{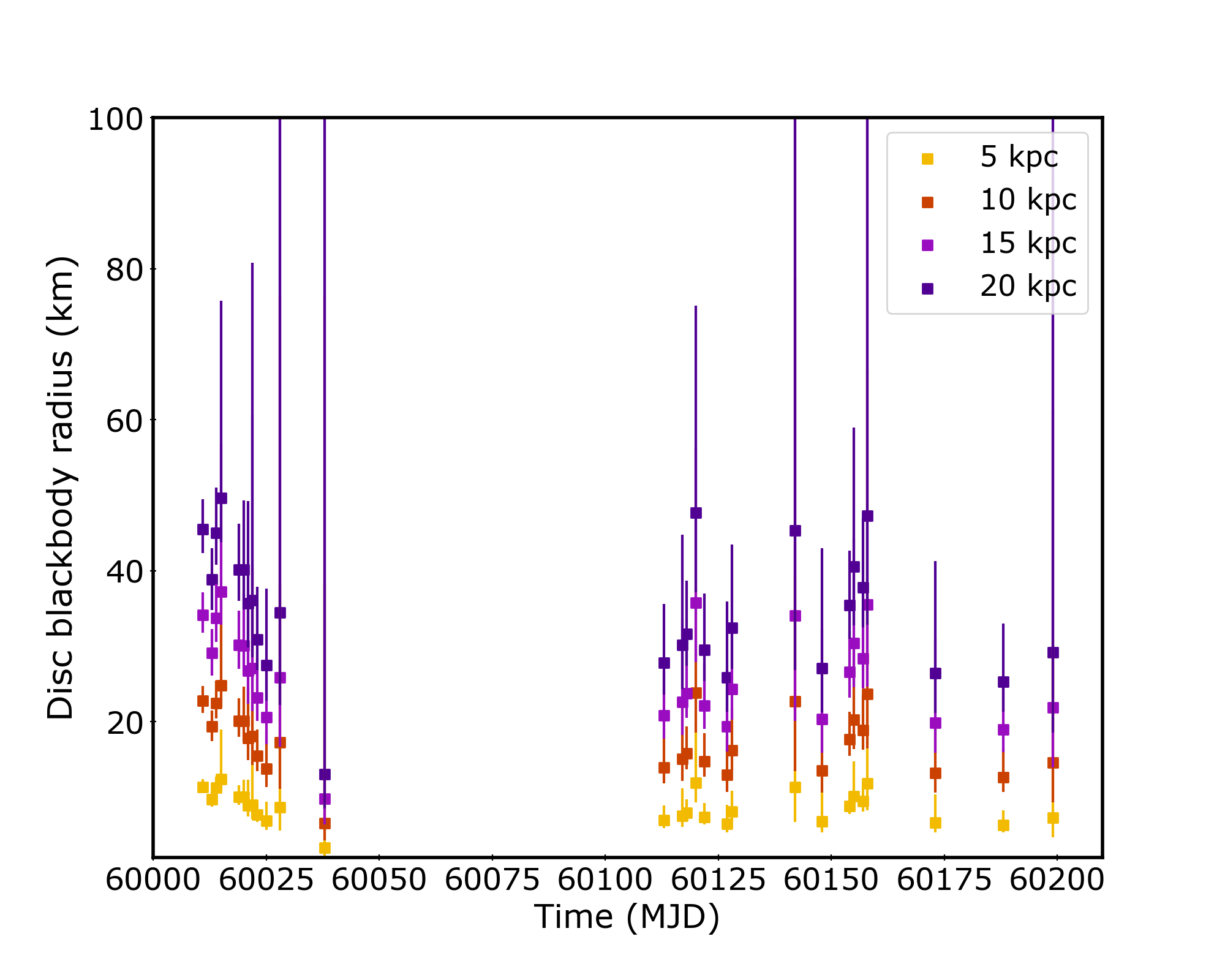} 
\vspace{-0.3cm}
\caption{\label{fig:radii} Estimates for the sizes of the black body emitting region (Model 1A, {\it Left}) and the disc (Model 1B, {\it Right}) as calculated from $K_{\rm bbody}$ and $K_{\rm disc}$, respectively.}
\end{figure*}

The best-fitting values for the \nicer and \swift-XRT spectra are listed in Tables\,\ref{tab:single-nicer-1a}, \ref{tab:single-nicer-1b} and \ref{tab:spec_swift}.  Figure\,\ref{fig:tower-plot-nicer} shows the temporal evolution of the spectral parameters and the unabsorbed flux derived in the 0.5--10\,keV energy band. The general trends observed in the spectral parameters remain consistent regardless of whether a \texttt{diskbb} or a \texttt{bbodyrad} component is used to describe the thermal emission. The thermal components were found statistically significant only in the brightest phases of the outburst.
A correlation between the X-ray flux and the $\Gamma$ index is evident in all fits, with the source showing a softening trend as it becomes brighter in X-rays (see also Figure\,\ref{fig:flux-gamma}). We can use the normalisations of the black body $K_{\rm bbody}$ and the disc black body $K_{\rm disc}$ components to obtain an estimate of the size of those regions. We have:

\begin{equation}
    { K_{\rm bbody}} = \bigg ( \frac{R_{\rm bb}}{D_{10 {\rm kpc}}} \bigg)^2 , \\
    { K_{\rm disc}} = \bigg ( \frac{R_{\rm in}}{D_{10 {\rm kpc}}} \bigg)^2 \cos\theta,
\end{equation}

\noindent where $R_{\rm bb}$ and $R_{\rm in}$ are expressed in km, $D_{10 {\rm kpc}}$ is the distance in units of 10~kpc, $\theta$ is the inclination angle of the system. We currently have no information on the distance of the source, so we calculated the radii for four possible values of the distances, as shown in Fig. \ref{fig:radii}. The inclination angle is also unknown, so that we kept it fixed at 60\deg. The obtained size of the black body emitting region, that is several tens of km for any tested distance value, would be more compatible with a boundary layer or a narrow disc-like region than, for example, a fraction of the NS surface. However, we caution that the lack of knowledge on the source distance and/or the inclination does not allow us to rule out completely the NS surface scenario. Given the practical equivalence between Models 1A and 1B and the rough indication of a disc-like emitting region, in the following sections we will mainly adopt Model 1B (\texttt{diskbb} as thermal component).


\subsubsection{Low-energy absorption feature} \label{sec:lowenfeature}
The first eight \nicer spectra show a hint of an absorption feature at $\sim 0.96$~keV, as visible in the \nicer and \nustar broadband spectrum in Figure \ref{fig:abs-features}. To test its origin and significance, we proceeded to merge the \nicer observations that have compatible spectral index and similar hardness ratio. We repeated the procedure for three groups of observations: 6203690101 to 6203690107, showing the highest hardness ratio; 6203690135 to 6203690141, showing the lowest hardness ratio; 6203690157 to 6203690169, during the reflaring phase. We used Model 1B for the first and third groups, and Model 0 for the second group, as described in Section \ref{sec:spectral}, with the addition of the \texttt{gaussian} component (with negative normalisation) to model the feature. 
Therefore, the models read as \texttt{TBabs(nthComp + diskbb + gaussian)} and \texttt{TBabs(nthComp + gaussian)} respectively. We found a rough estimate of the significance of the feature as the ratio of its normalisation over the error at 1$\sigma$ of the same parameter. The second group did not show any significant feature, while in the first and third groups we noticed a feature with significance of $\sim 4.2 \sigma$ in the softest spectra, and $\sim 2.6 \sigma$ during the reflaring phase. In the first group, the feature is centred at 0.94$^{+0.02}_{-0.07}$\,keV, with width $\sigma_{\rm line}$=0.07$^{+0.03}_{-0.03}$\,keV and equivalent width $\sigma_{\rm eq}$=22.6$^{+0.8}_{-0.9}$~eV. During the reflaring phase, we find $E_{\rm line} = 0.93^{+0.04}_{-0.05}$~keV, $\sigma_{\rm line}< 0.12$~keV, and $\sigma_{\rm eq} < 25.8$~eV.

To further investigate the presence of this spectral line, we inspected the \swift-XRT spectra as well. We noted that the individual spectra do not show evidence of the feature. Therefore, we decided to combine all the spectra corresponding to the WT-mode observations. We fit the combined spectrum with Model 0 ($\chi^2_\nu$=1.3 for 56 degrees of freedom (dof)) and structured residuals around 1\,keV were visible. The inclusion of the \texttt{gaussian} component improved the fit yielding $\chi^2_\nu$=1.1 for 53 dof. The line parameters are the following: $E_{\rm line}$=0.96$^{+0.04}_{-0.29}$\,keV, $\sigma_{\rm line}<$0.41\,keV and $\sigma_{\rm eq}$=28$\pm$17\,eV, with a significance of $\sim$2$\sigma$. We also merged the PC-mode observations with consistent flux levels and did not find any hint of the absorption feature. 

Considering the hint for the presence of the feature in both \nicer and \swift spectra, we decided to keep the component in the modelling of the broadband spectrum (see Section \ref{sec:broadband}). In order to make use of a more sophisticated model for the absorption feature, we verified that the modelling with the \texttt{gaussian} component is equivalent to the modelling with the \texttt{gabs} component, and eventually adopted the second option for our final version.

\subsubsection{The \nicer+\nustar broadband spectrum}
\label{sec:broadband}
We fitted the \nicer (OBSID: 6203690104) and \nustar spectra extracted from the simultaneous observations performed on March 10, 2023.
We analysed the \nicer observation in the 0.5--5.9~keV energy range (exposure time=7.6\,ks), and the \nustar data in the 3.0--60.0~keV energy range (FPMA exposure time=29.0\,ks and FPMB exposure time=28.8\,ks). Both energy ranges were chosen to limit our analysis to the intervals of the spectra that are dominated by the source emission. In all the following models, we kept all the parameters, including the photon index, linked between the spectra, and we adopted a \texttt{constant} component (fixed at 1 for the \nicer spectrum) to take into account the cross-calibration differences between the instruments.

\begin{figure*}
    \centering
    \begin{subfigure}{0.48\textwidth}
        \includegraphics[width=\textwidth]{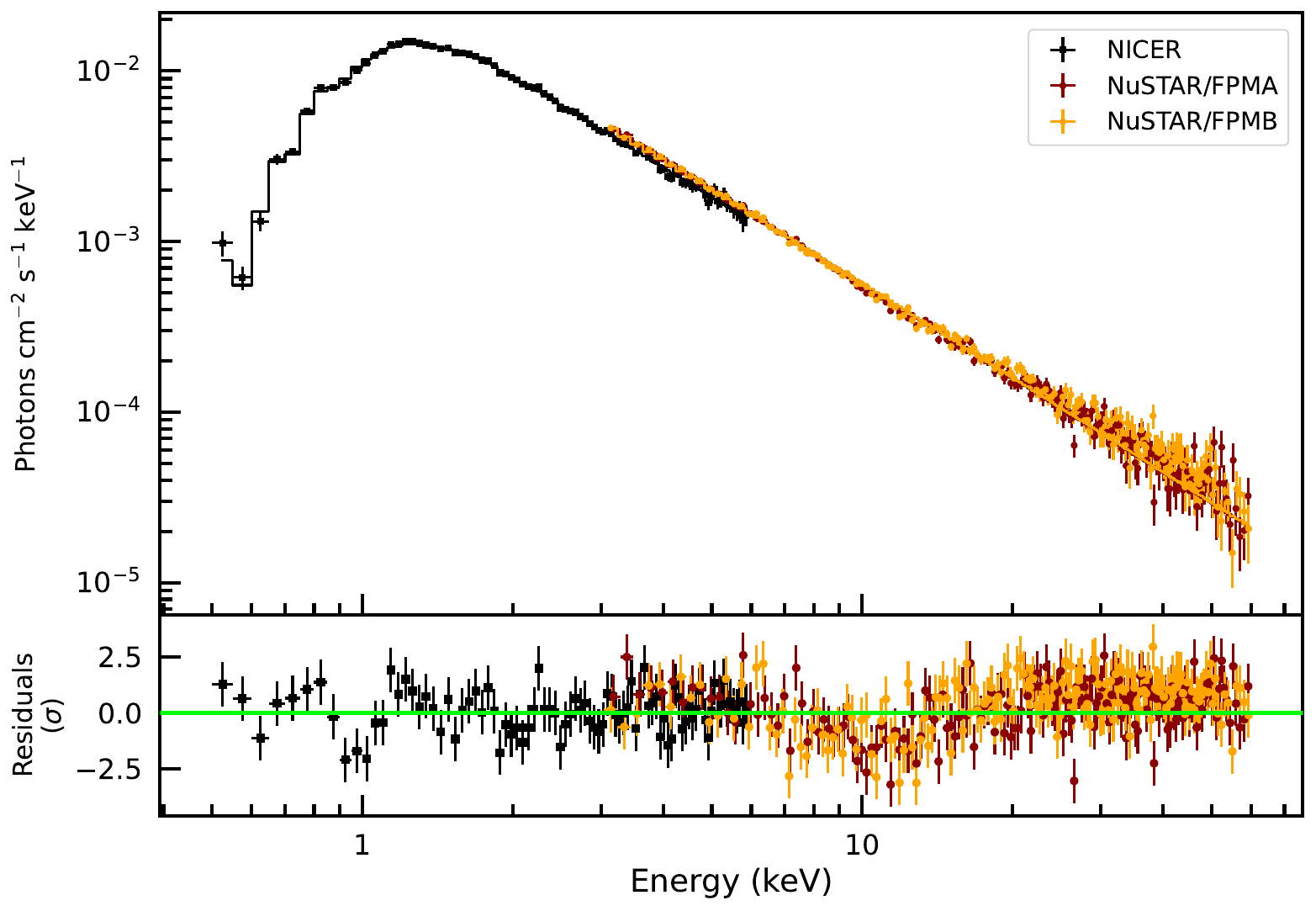}
        \caption{Model 1B without the absorption features}
        \label{fig:abs-features}
    \end{subfigure}
    \begin{subfigure}{0.48\textwidth}
        \includegraphics[width=\textwidth]{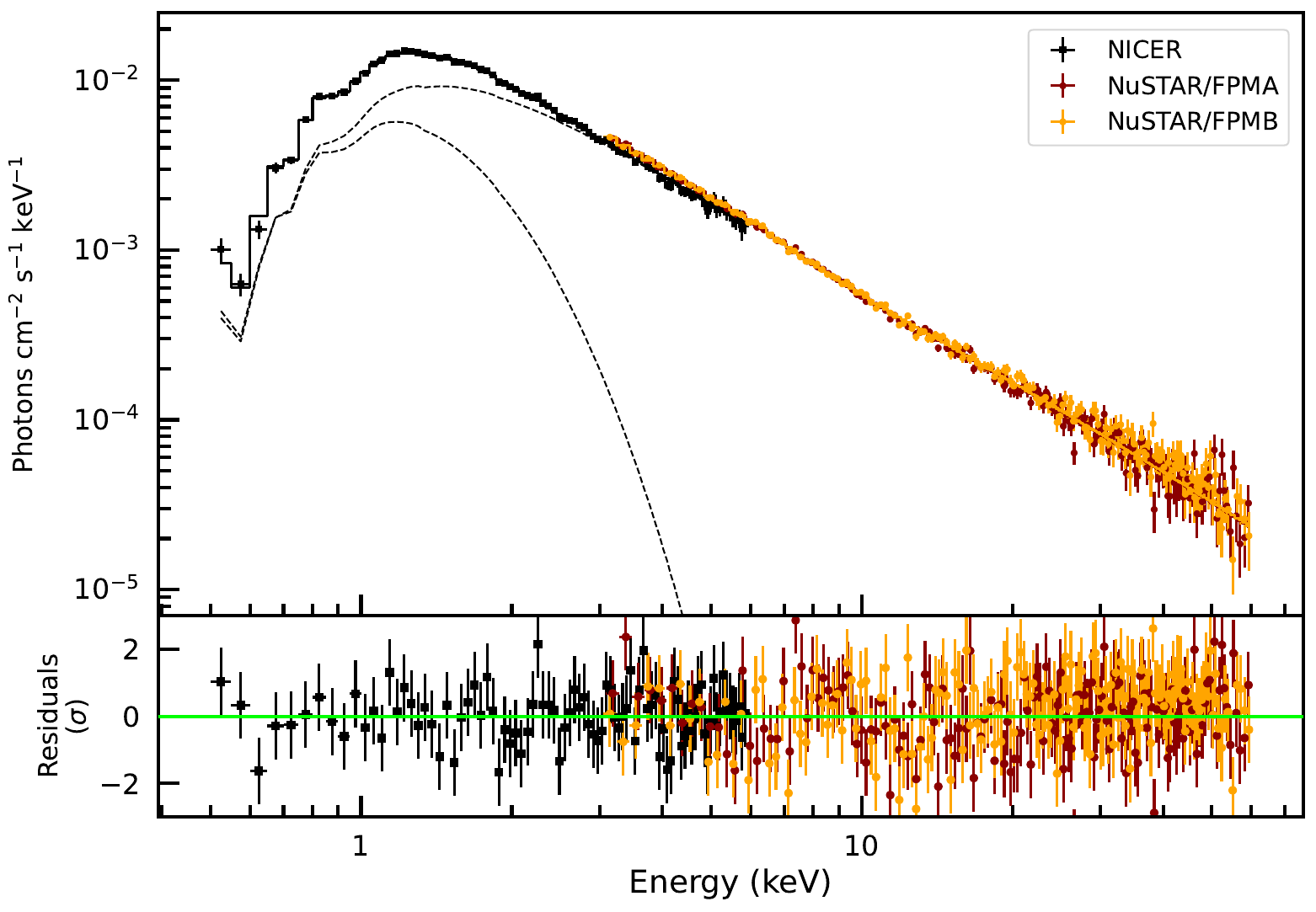}
       \caption{Model 1B}
       \label{fig:nthcomp}
    \end{subfigure}
    \begin{subfigure}{0.48\textwidth}
        \includegraphics[width=\textwidth]{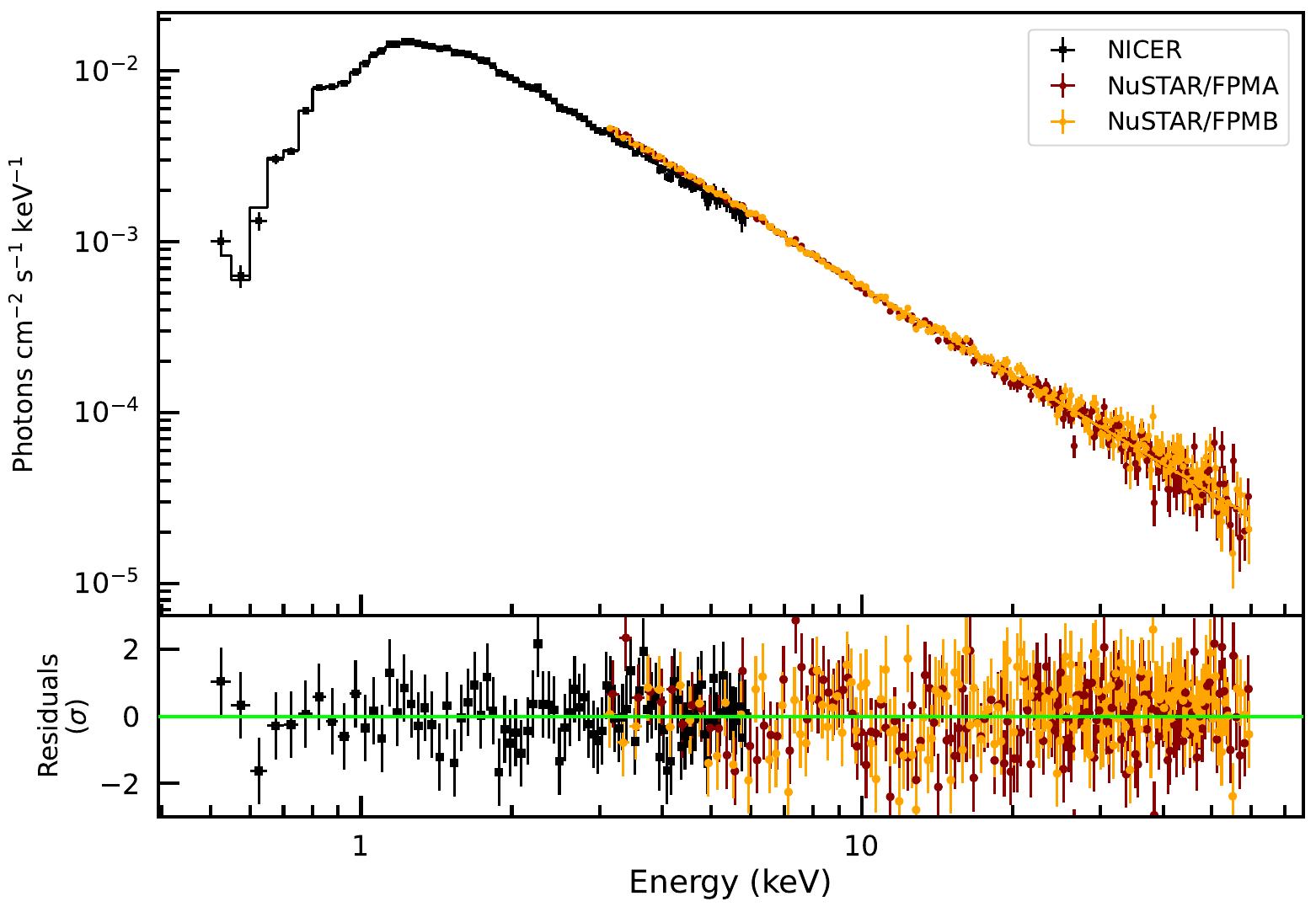}
       \caption{Model 2}
       \label{fig:simpl}
    \end{subfigure}
    \begin{subfigure}{0.48\textwidth}
        \includegraphics[width=\textwidth]{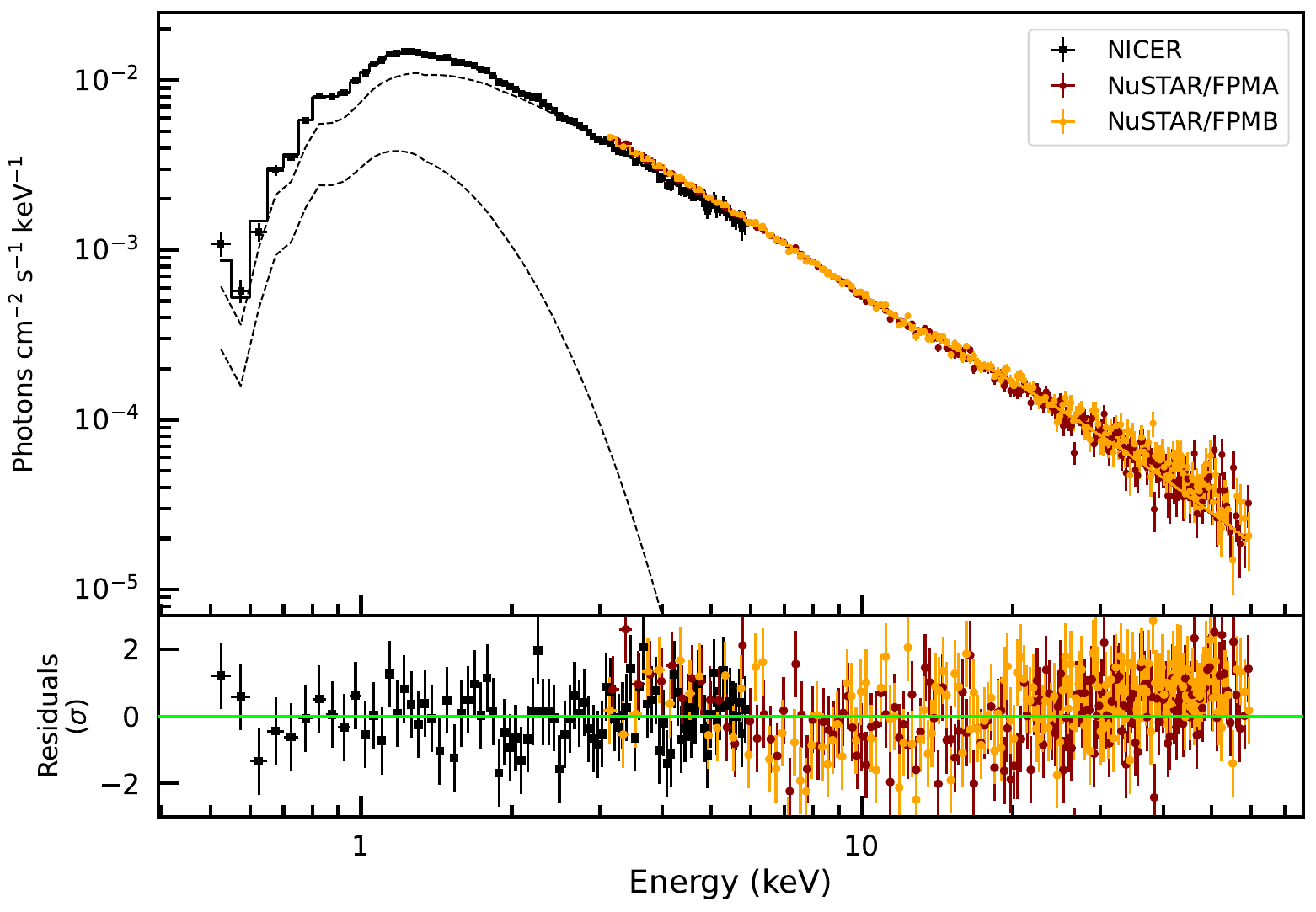}
       \caption{Model 3}
       \label{fig:relxillcp}
    \end{subfigure}
    \caption{Unfolded broadband spectrum (upper panel) and fit residuals (lower panel) extracted from \nicer (black)+\nustar/FPMA (red)+\nustar/FPMB (orange) data over the 0.5--60\,keV energy range. Panel (a) shows the fit without the inclusion of the absorption features (\texttt{TBabs(nthComp + diskbb)}).
    Panels (b) and (c) show the fit with Model 1B (\texttt{TBabs$\times$gabs$\times$smedge(nthComp + diskbb)}) and Model 2  (\texttt{TBabs$\times$gabs$\times$smedge$\times$simpl$\times$diskbb}), respectively. Panel (d) shows the fit with Model 3 (\texttt{TBabs$\times$gabs(diskbb + relxillCp)}).} 
    \label{fig:broadband}
\end{figure*}

Firstly, we tested Model 1B, with the addition of the \texttt{gabs} component as reported in Section \ref{sec:lowenfeature}, whose centroid was kept fixed at an energy of 0.97~keV. The electron temperature $kT_{\rm e}$ could not be constrained even in the broadband spectrum, therefore we kept it fixed at 100~keV, beyond the range covered by \nustar. We linked the disc temperature with the seed temperature of the Comptonised component and assumed that seed photons originate from within the accretion disc (i.e., we set the \texttt{inp\_type} parameter to 1). As seen in Figure \ref{fig:abs-features}, we noticed a more pronounced and rather large absorption feature at $\sim 10$~keV, likely an Fe K-edge, which we fitted with an additional smeared edge (\texttt{smedge}) component. We limited the cut energy to values higher than 7~keV and fixed the smearing width to 2~keV. The new model 1B for the broadband spectrum is therefore \texttt{TBabs$\times$gabs$\times$smedge(nthComp + diskbb)}.
We also tried an alternative model replacing the \texttt{nthComp} model with a simpler convolution model (\texttt{simpl}). We noticed the new version gives higher values for the normalisation of the Comptonised component. We call this version Model 2: \texttt{TBabs$\times$gabs$\times$smedge$\times$simpl$\times$diskbb}. After fitting Model 2 to the data, we still noticed a slight excess in the residuals beyond 10~keV that could be due to the Compton hump and, together with the Fe K-edge, shows a hint for the presence of reflection processes, even if there is no clear evidence of an iron line. Therefore, we decided to add a reflection component, \texttt{relxillCp} \citep{GarciaDauser2014}, which substitutes the previous Comptonisation and smeared edge components (i.e., Model 3: \texttt{TBabs$\times$gabs(diskbb+relxillCp)}). We kept the outer radius fixed at the default value of 400~$R_{\rm g}$ (where $R_{\rm g} = GM/c^2$ is the gravitational radius), linked to the break radius (defined as the radius at which the disc emissivity may change), and the inner radius was allowed to vary. We also linked the emissivity indices and kept them at the default value of 3 \citep[standard Shakura-Sunyaev disc,][]{ShakuraSunyaev1973}. The electron temperature $kT_{\rm e}$ was frozen at 100~keV since we could not constrain it within our energy range. We also fixed the density of the disc and the iron abundance ($n_{\rm e} = 10^{19}$~cm$^{-3}$, 
$A_{\rm Fe} = 1$ solar abundances from the default value). The ionisation parameter was left free to vary. The redshift was set to 0 (Galactic source). The adimensional spin was set to 0.99 (corresponding to the hypothesis of a maximally rotating BH), though Model 3 appears to be insensitive to its value. Since the fit could not constrain the inclination, we kept it fixed at 60\deg. 


Table \ref{tab:broadband} reports the best-fit values for the broadband spectrum across all the models we investigated.
The spectrum is already well-modelled using a new version of Model 1B (Fig. \ref{fig:nthcomp}), with a Comptonised component with a spectral index $\Gamma \sim 1.8$, a disc temperature at inner radius of $\sim 0.4$~keV and a smeared Fe K-edge with cut energy of $\sim 7.5$~keV ($\chi^2_\nu=1.07$ for 405 dof).
In the case of Model 2, the spectral index of the Comptonised component and the disc temperature are consistent with those in Model 1B.
The fraction of up-scattered photons is $\sim $0.45. The absorption edge has a threshold energy of $\lesssim 7.8$~keV. In the 0.5--60~keV energy band, the unabsorbed flux is $\sim 4 \times 10^{-10}$~\unitF. We obtained $\chi^2_\nu=1.05$ for 405 dof, showing no significant statistical improvement over Model 1B. However, the normalisation of the disc component increases substantially from $\sim 163$ for Model 1B, to $\sim 345$ for Model 2. Model 3 yields similar values for the spectral index and the disc temperature. 
We can constrain the logarithm of the ionisation parameter of the disc matter to $\sim 3.7$. We can only determine a 3$\sigma$ upper limit on the disc inner radius of 11.4~$R_{\rm g}$. The unabsorbed flux in the 0.5--60~keV energy band remains $\sim 4 \times 10^{-10}$~\unitF. 


\begin{table*}
    \centering
    \caption{Best-fit values of the \nicer+\nustar\ broadband spectrum. The fluxes are calculated in the 0.5--60~keV energy band. Models used: Model 1B: \texttt{TBabs$\times$gabs$\times$smedge(nthComp + diskbb)} , Model 2:\texttt{TBabs$\times$gabs$\times$smedge$\times$simpl$\times$diskbb} , Model 3: \texttt{TBabs$\times$gabs(diskbb + relxillCp)}.\label{tab:broadband}}
    \begin{tabular}{ll|ccc}
    \hline \hline
      Component & Parameter & Model 1B & Model 2 & Model 3\\
      \hline 
      \texttt{TBabs} & $N_{\rm H}$ ($\times 10^{22}$~cm$^{-2}$) 
      &0.84$^{+0.03}_{-0.03}$ & 0.84$^{+0.02}_{-0.02}$ & 0.90$^{+0.04}_{-0.01}$ \\
      \hline
       \multirow{3}{*}{\texttt{gabs}}
      & E (keV) & 0.97$^*$ & 0.97$^*$ & 0.97$^*$ \\
      & $\sigma$ (keV) & 0.06$^{+0.03}_{-0.04}$ & 0.05$^{+0.04}_{-0.04}$ & 0.07$^{+0.04}_{-0.04}$ \\
      & Strength (keV) & 0.015$^{+0.009}_{-0.007}$ & 0.015$^{+0.008}_{-0.007}$ & 0.021$^{+0.011}_{-0.009}$ \\ \hline
      \multirow{2}{*}{\texttt{diskbb}} & $kT_{\rm disc}$ (keV) & 0.38$^{+0.03}_{-0.03}$ & 0.37$^{+0.02}_{-0.02}$ & 0.34$^{+0.03}_{-0.04}$ \\
       & $K_{\rm disc}$ & 163$^{+66}_{-37}$ & 345$^{+122}_{-80}$ & 177$^{+226}_{-59}$ \\ 
       \hline
    \texttt{nthComp} & $\Gamma$ & 1.79$^{+0.01}_{-0.01}$ & - & - \\
    \hline
     \multirow{2}{*}{\texttt{simpl}} & Scattered frac. & - & $0.45^{+0.02}_{-0.06}$ & - \\ 
                                     & $\Gamma$ & - & 1.79$^{+0.01}_{-0.02}$ & - \\
     \hline
     
      \multirow{2}{*}{\texttt{smedge}}
      & $E_{\rm edge}$ (keV) & 7.5$^{+0.5}_{-0.5}$ & $\lesssim 7.8$ & - \\ 
      & Absorption depth & 0.26$^{+0.04}_{-0.04}$ & 0.25$^{+0.04}_{-0.04}$ & - \\ 
      & Index for photo-electric cross-section & -2.67$^*$ & -2.67$^*$ & - \\ 
      & Smearing width (keV) & 2$^*$ & 2$^*$ & - \\ \hline
      \multirow{5}{*}{\texttt{relxillcp}} & $\Gamma$ & - & - & 1.72$^{+0.01}_{-0.01}$ \\
      & $R_{\rm in}$ (R$_g$) & - & - & < 11.4 \\
      & $A_{\rm Fe}$ (A$_{\odot}$) & - & - & 1$^*$ \\
      & log X$_i$ & - & - & 3.69$^{+0.03}_{-0.02}$\\
      & log ($n_{\rm e}$/cm$^{-3}$) & - & - & 19$^*$ \\
      & Refl. frac. & - & - & > 3 \\ \hline
      Observed Flux ($\times 10^{-10}$ \unitF) & & 3.29$^{+0.13}_{-0.03}$ & 3.31$^{+0.14}_{-0.34}$ & 3.21$^{+0.05}_{-0.28}$ \\
      Unabsorbed Flux ($\times 10^{-10}$ \unitF) & & 3.78$^{+0.28}_{-0.21}$ & 4.10$^{+0.13}_{-0.11}$ & 4.06$^{+0.05}_{-0.04}$ \\
      $\chi^2$ (dof) & & 435.02(405) & 426.71(405) & 442.69(404) \\
      \hline
    \footnotesize{$^*$ Fixed parameter in the fit.}
    \end{tabular}
\end{table*}

\subsection{Timing analysis}\label{sec:timing}

For the timing analysis, the photon arrival times of the source event files were referred to the Solar System barycentre using the \texttt{barycorr} tool, the latest calibration files, the ephemeris DE\,405 and the coordinates derived in Section\,\ref{sec:position}.

\begin{figure}
\centering
\includegraphics[width=1.\columnwidth]{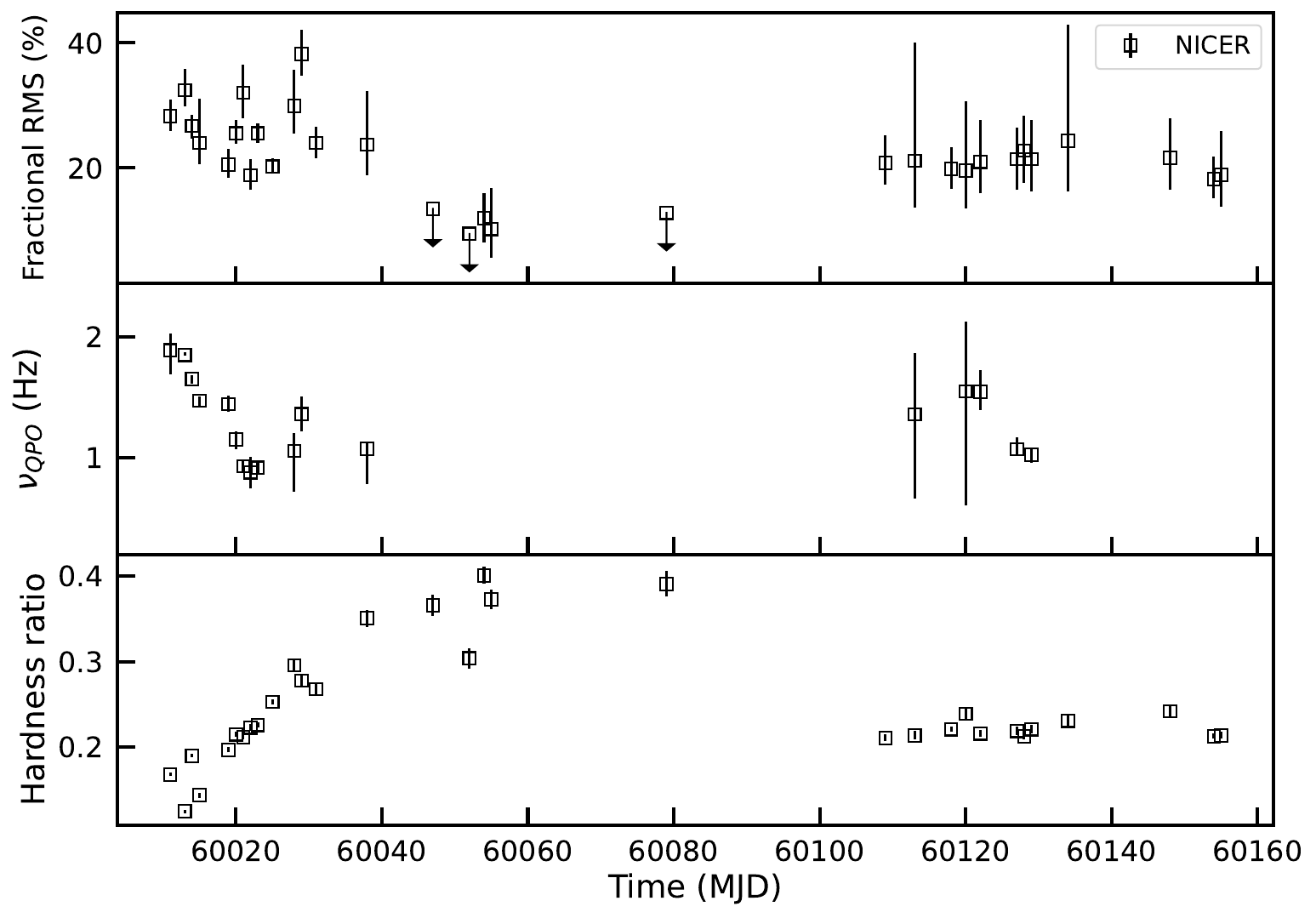}
\includegraphics[width=1.\columnwidth]{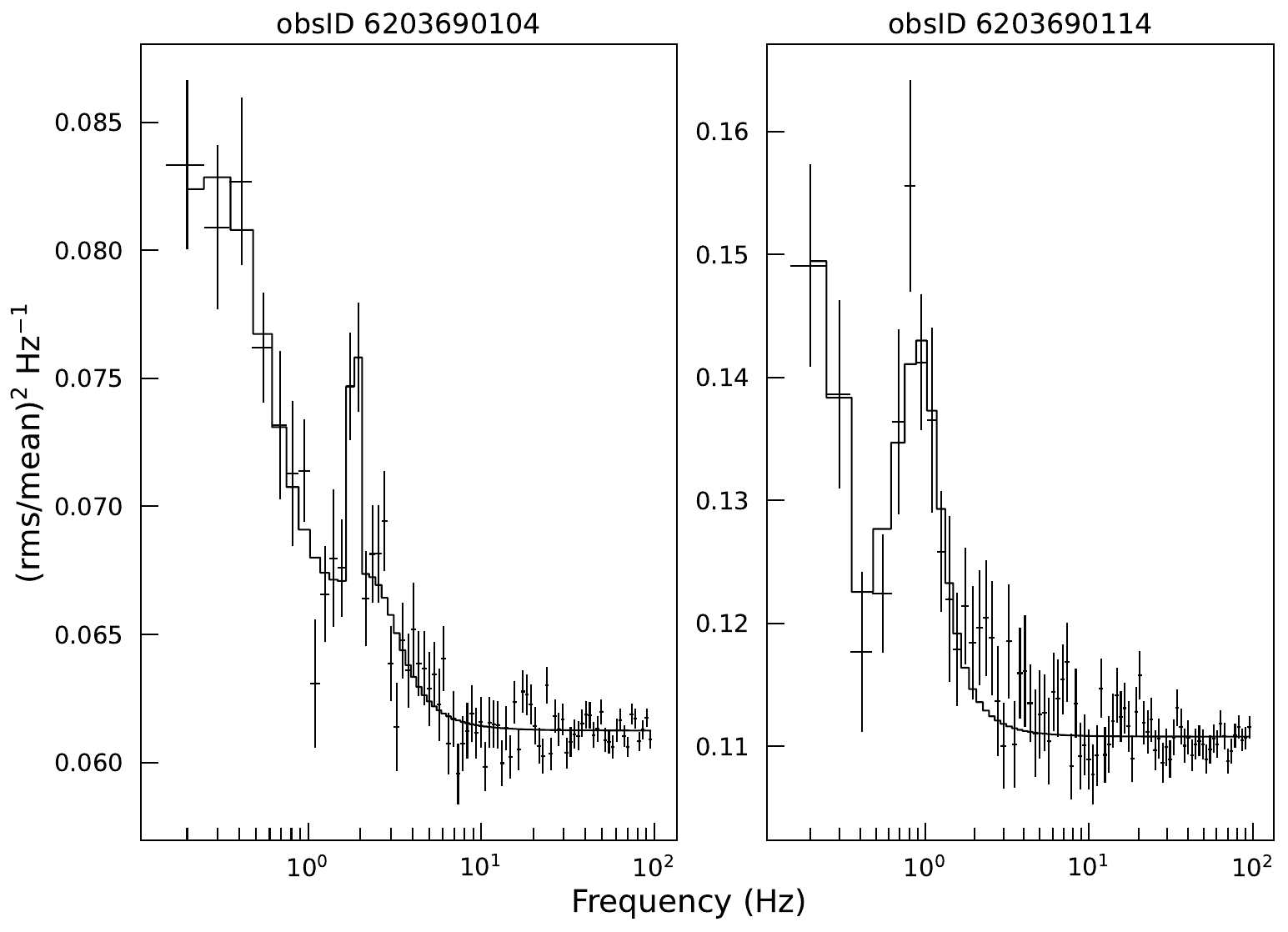}
\includegraphics[width=1.\columnwidth]{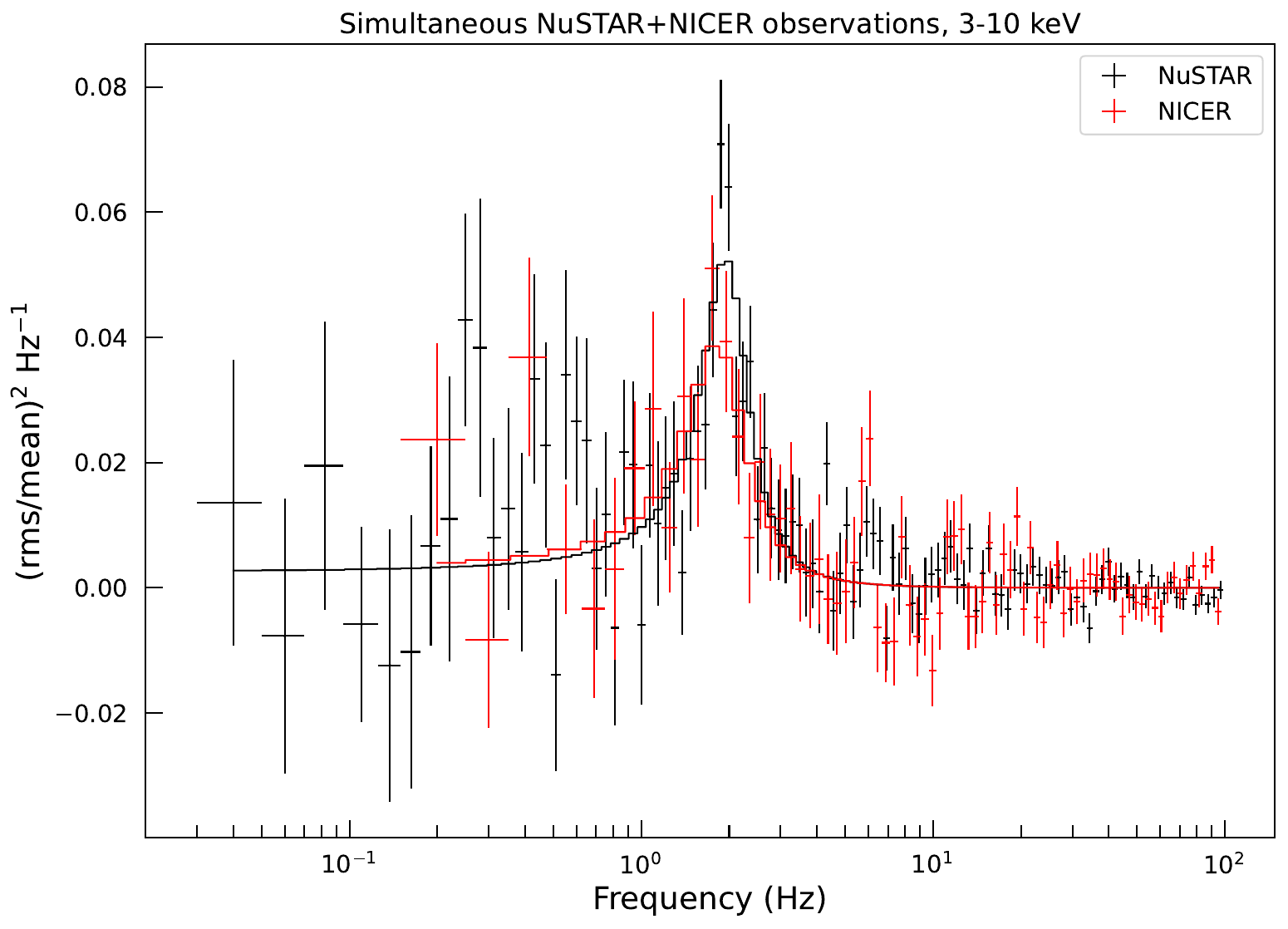}
\vspace{-0.3cm}
\caption{\label{fig:pds} {\it Top}: temporal evolution of the RMS integrated over all frequencies (top panel) and QPO central frequency (medium panel) estimated from the {\bf 0.3--12\,keV} \nicer PDS. In the bottom panel we present the temporal evolution of the hardness ratio, as defined in Sec.\,\ref{sec:HID}. {\it Middle}: Two {\bf 0.3--12\,keV} \nicer PDS showing the QPO, in the case of the highest (left, OBSID 6203690104) and lowest (right, OBSID 6203690114) central frequency $\nu_{\rm QPO}$. The Poisson noise was not subtracted. {\it Bottom}: The PDS of the \nustar\ (black) and \nicer (red) simultaneous observations in the common energy, 3--10\,keV. The Poisson noise was subtracted for plotting purpose. The solid lines represent the best-fit models.}
\end{figure}

In order to probe the source short-term X-ray variability, we computed the PDS by normalizing the power to the squared fractional root mean square (RMS) from \nustar and \nicer event files using the 3--79\,keV and 0.3--12\,keV energy range, respectively, integrated over all the frequencies. We applied the dead-time correction on the \nustar PDS using the Fourier Amplitude Difference (FAD) method \citep{bachetti2018} as implemented in {\tt Stingray} \citep{Huppenkothen2019,Bachetti2024} and then extracted PDS averaging over 50-s long segments with time bin of 0.5\,ms. For \nicer we used only observations with more than 5000 photons, and split them into stretches with a duration of 10\,s and a timing resolution of 0.6\,ms. The PDS created for each segments were averaged to produce an average PDS per observation. 
For each PDS, we estimated the fractional RMS over the entire frequency band by modelling the PDS in {\sc Xspec} with a combination of Lorentzian functions and a constant to take into account the Poisson noise contribution. The RMS of the 3--79\,keV \nustar PDS was 35$\pm$4\%, while we show the temporal evolution of the RMS for the 0.3--12\,keV \nicer PDS in Figure\,\ref{fig:pds} (top panel). During the first 40 days of the monitoring campaign, the RMS did not show any particular trend, oscillating around an average of $\sim25\%$. It reached its minimum ($\sim10\%$) when the flux was also at its lowest value. Afterwards, the RMS increased again and attained a constant value of $\sim20\%$ during the rebrightening phase. We detected a QPO in some of the \nicer observations; two examples of PDS exhibiting the timing feature are displayed in Fig.\,\ref{fig:pds} (middle panel). By using a Lorentzian component to model the QPO, we studied the temporal evolution of its best-fit centroid frequency $\nu_{\rm QPO}$, which decreased from $\sim$1.9 to 0.9\,Hz in the first $\sim20$ days of our campaign, as shown in Fig.\,\ref{fig:pds}. In addition, an anti-correlation was found between $\nu_{\rm QPO}$ and the hardness ratio: when the former decayed, the latter increased until the rebrightening, when both quantities were constant. For the simultaneous \nicer and \nustar observations, we extracted the PDS in the common energy band, 3--10\,keV (see Fig.\,\ref{fig:pds}, bottom panel). The best-fit $\nu_{\rm QPO}$ was 1.8$\pm$0.2\,Hz for \nicer data and 1.9$\pm$0.1\,Hz for \nustar data, while the full width at half maximum was equal to 1.1$^{+0.9}_{-0.5}$\,Hz and 0.9$\pm$0.3\,Hz for the \nicer and \nustar PDS, respectively.


We also searched for periodic signals in the \nustar observation, as well as in the first four \nicer observations individually (i.e., those providing the largest counting statistics) using Fourier-domain acceleration search techniques. For each data set, we considered the whole energy range as well as distinct energy bands. We used the \texttt{accelsearch} pipeline from the \texttt{PRESTO}\footnote{\url{https://github.com/scottransom/presto}} pulsar timing software package \citep{ransom02} to search for signals over the frequency range of 1--2000\,Hz, summing up to 2, 4 and 8 harmonics. To account for potential power drifts in the Fourier domain, we allowed the powers of signals to drift by up to 200 frequency bins. Additionally, we tested the case where the powers of signals could drift by up to 600 frequency derivative bins (known as `jerk' search; \citealt{andersen18}). This analysis was conducted on the entire observation dataset as well as on data chunks of 300\,s.  The identified candidate periodicities were sifted so as to reject less significant, duplicate, and/or harmonically related candidates.
No statistically significant signal was detected.

\section{Discussion}
We analysed the 2023 outburst of the X-ray transient \source with \nicer, \nustar, and \swift data from March to October 2023. We conducted a spectral analysis of all the single observations and of a broadband \nicer+\nustar spectrum. 
No NS observational signatures, such as type-I X-ray bursts or pulsations, have been detected to rule out the possibility of the presence of a BH. We also performed a timing analysis of the \nicer and \nustar observations, deriving the value of the RMS to help determine the state of the system, and studying possible QPO components.

\subsection{Outburst history: evolution of the spectral parameters}\label{ss:disc-history}
Since the discovery in February 2023, \source was monitored for about 7 months, from March 3 to October 2, 2023, until it became no longer visible by both \swift\ and \nicer. The source displayed a structured light curve with one main outburst observed mainly during its decay phase. The unabsorbed flux in the 0.5--10~keV energy range decreased from $\sim2.5\times10^{-10}$~\unitF down to $5.5\times10^{-12}$~\unitF over three months, followed by a fast rebrightening and a series of flux variations between $\sim3.5\times10^{-11}$ and $7\times10^{-11}$~\unitF (see Fig.\,\ref{fig:tower-plot-nicer}, top panel). During the main outburst, both \nicer\ and \swift-XRT spectra were dominated by a non-thermal component, well-described by a Comptonisation spectrum with $\Gamma\sim$1.5--2.2. An additional thermal component, either a black body or a disc black body, is required in \nicer\ data for the first 30 days of the monitored period, with a slight decrease in temperature from $\sim$0.4 keV to $\sim$0.3 keV. This thermal component later became not significant. 
The non-thermal nature of the spectrum suggests that during our monitoring campaign \source was in a hard or intermediate state; this classification is corroborated by the RMS values found during the campaign \citep[e.g.,][]{MunozDarias2011, MunozDarias2014}. It is noteworthy that the follow-up X-ray observations were only triggered about a month after the initial \textit{MAXI} detection, due to a delay in the announcement \citep{Atel15929}. We checked whether a possible transition to a soft state could be detected in the \textit{MAXI} data during the one-month delay. However, we did not find any clear softening in the light curve or in the hardness ratio. \source could be one of the transient X-ray binaries that underwent a failed-transition outburst \citep[e.g.][]{Alabarta2021}. 
Around MJD 60100, the source rebrightened, undergoing a second longer outburst that lasted until the end of the 2023 visibility window in October. These rebrightening or `echo outbursts' \citep[e.g.][]{Zhang2019} have been observed often in both BH \citep[e.g.,][]{Cuneo2020} and NS \citep[e.g.,][]{Patruno2016}. The rebrightening in \source\ never reached the flux level of the first main outburst, achieving a flux of $\sim$1$\times$10$^{-10}$~\unitF at peak, corresponding to 60\% of the peak flux of the first activity phase of the outburst. Interestingly, the pattern followed by the source in this second activity phase is rather erratic, with the flux swinging around an average level of (5--6)$\times$10$^{-11}$~\unitF. We tried to model all the observations using both Model 0 and 1A/1B, and interestingly found that only a subset of the spectra required the inclusion of a thermal component in the model, namely those collected during periods of flux increase. 
This behaviour is reminiscent of what observed in the BH MAXI J1348--630 during its 2019 reflare \citep{Dai2023}. These authors, in particular, suggested that the disc started an ingress trend once the system exceeded a `critical' luminosity of $\sim$2.5$\times$10$^{36}$~\unitL. 
Assuming a similar critical threshold for our system, we can obtain a possible value for the distance for the source of about 19 kpc. We note that such a high distance could explain the radio non-detection of the source (see Section \ref{sec:mw}); however, given the many assumptions made to obtain this distance value, we caution that this has to be taken more as a hint than an actual measurement. 




\begin{figure}
\centering
\includegraphics[width=1.\columnwidth]{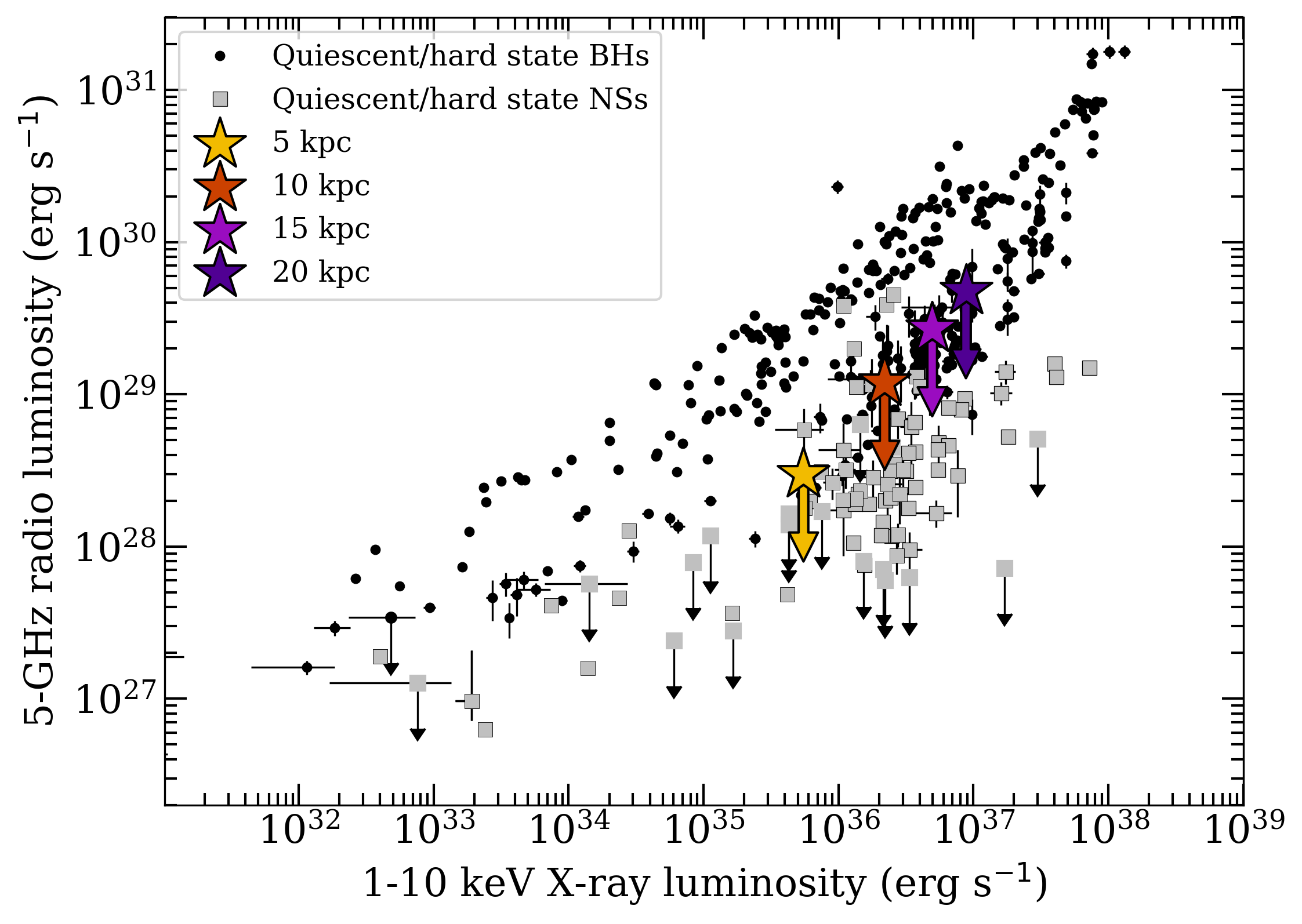} 
\vspace{-0.3cm}
\caption{\label{fig:lrlx} Radio:X-ray luminosity diagram of quiescent/hard state BH transients (black circles) and NS (grey squares) LMXBs, plotted using the data repository by \citet{Bahramian2018}. In the plot, the X-ray luminosity of \source and the upper limit on the radio luminosity obtained by \citet{Atel15939} are plotted with stars, colour-coded for different distances.}
\end{figure}

\subsection{Broadband spectrum and spectral features}
The continuum of the broadband spectrum is described by a model composed of a multi-colour disc black-body and a Comptonised emission component. We could not constrain the electronic temperature within our energy range, hinting at the fact that the high-energy cutoff might extend beyond 60~keV of the \nustar energy range. Such values for the high-energy cutoff are typical of BH systems, while for NS systems, the cutoff temperature is found within $\sim 15-25$~keV \citep{Burke2017}. The spectrum showed a dip at around $\sim10$~keV, which we interpreted as a potential smeared Fe K-edge. With the addition of a smeared edge component, we obtained a cut energy of $\sim7.5$~keV, compatible with Fe K-shell recombination processes. Nonetheless, we find no evidence for the presence of an iron emission line. 
However, due to the evidence of a potential Fe K-edge, we included a reflection component to the model. 
The new model is still statistically equivalent to the previous one (see Table\,\ref{tab:broadband}). The reflection component is not very strong and we could obtain only a lower limit on the ionisation parameter of ${\rm log} X_i\gtrsim 3.7$, corresponding to highly ionised matter and, thus,  
aligning with the absence of a detectable iron line at $\sim 6.4$~keV \citep[for a description of the model, see][]{GarciaDauser2014}. We could derive an upper limit for the inner radius of the accretion disc at $\sim 11.4$~$R_{\rm g}$, but the adimensional spin parameter could not be properly determined. 
This does not exclude that the disc might extend to lower radii, as the radius derived from the reflection component corresponds to the radius at which recombination processes are taking place significantly enough to be detected, and does not necessarily have to match the inner radius of the disc \citep{Krolik2002}.
A more meaningful comparison could be done with the disc inner radius estimated from the thermal disc component of the model, but our derivation is currently affected by the lack of information on the inclination angle and on the source distance. We can only illustrate the dependence of the apparent inner radius on the unknown parameters. From the reflection component, we obtain an upper limit of $11.4 R_g \simeq 17 (M/M_\odot)$~km, corresponding to $\sim 24$~km for a mass of $1.4 M_\odot$ of the prototypical NS, and $\sim 170$~km for a  black hole of mass $10 M_\odot$. Referring to the values reported in Table\,\ref{tab:broadband}, while for Model 1B the normalisation is consistent with that of Model 3 (see Fig. \ref{fig:radii}), for Model 2 we obtain a normalisation of $\sim 345$, corresponding to an inner radius of $\sim$13--39 km for an inclination angle of 60\deg and a distance D in the range 5 to 15 kpc.



We detected an absorption feature at $\sim 0.96$~keV, both in the \nicer spectra and in the merged \swift-XRT WT spectra. The feature has a significance of $\sim 4 \sigma$ in the softest \nicer spectra, and of $\sim 2 \sigma$ in the \swift and \nicer spectra collected during the reflaring phase. While we cannot entirely rule out the possibility of an instrumental origin for this feature, its detection in both \nicer and \swift, although with low significance, data suggests it is not an artifact. A similar absorption feature at $\sim 1$~keV was recently detected in \nicer \citep{DelSanto2023} and \XMM (Del Santo et al., in prep.) observations of the BH candidate MAXI J1810--222. The authors explained the spectral feature as possible evidence of ultrafast outflows from the system.


\subsection{On the nature of the source}\label{sec:mw}
The nature of the accretor in \source remains unclear. Type-I X-ray bursts or pulsations, incontrovertible signatures of the NS nature of the primary star, have not been detected in \nustar, \swift-XRT or \nicer data (Section \ref{sec:timing}). However, the absence of these features does not rule out the presence of an NS, since the majority of NS LMXBs are not X-ray pulsators and type-I X-ray bursts can easily be missed \citep{DiSalvo2023}. In this context, X-ray spectral and timing properties can give some clues, though they are not conclusive. During the hard/intermediate state, \source shows a relatively hard spectrum with $\Gamma \lesssim$1.7 and high RMS values (never below 10\%), which are typically found in BH systems in hard/intermediate state \citep[see, e.g.,][]{MunozDarias2011}, though they are sometimes displayed also by NS systems \citep[see, e.g.,][]{MunozDarias2014}. 

A QPO is present at the beginning of the decay of the first activity phase with a central frequency decreasing from $\sim$1.9 to 0.9\,Hz. The QPO also appears in a few observations during the rebrightening. Common in both BH and NS LMXBs, QPOs have been observed across a wide range of frequencies and have been classified into different types. The feature in the PDS of \source is characterised by a narrow peak, as observed for Type-C and B QPOs in BH binaries \citep[e.g.,][]{Motta2015}.
Type-C QPOs are accompanied by a strong flat-top noise component, while, Type-B QPOs generally appear in the PDS coincident with weak red noise. For the former, the central frequency is tightly correlated with the spectral state, rising from a few mHz in the hard state to $\sim$10\,Hz in the intermediate states. For the latter, it 
is usually in the 5--8\,Hz range, but in a few instances type-B QPOs were also found at 1--3\,Hz. The archetypal BH binary GX\,339--4 shows type-B QPOs at low frequencies \citep[i.e. below $\sim$3\,Hz, see][]{motta2011} during intermediate states, and more specifically during the low luminosity soft-to-hard transition, which might be the case of \source. However, GX\,339--4 showed softer energy spectra and lower RMS values in correspondence with these low-frequency type-B QPOs than what we observed for \source. Also type-C QPOs at $\sim$1--2\,Hz are found in GX\,339--4 when the RMS reaches values as high as those estimated for this new transient \citep{motta2011}.
For NS systems, the QPO phenomenology is richer. 
If the power spectral feature of \source were a normal branch oscillation, the centroid frequency would not match the expected values ($\sim$5--6\,Hz). However, there are quasi-periodic variability phenomena in NS LMXBs that do not have an equivalent in BH LMXBs. For example, QPOs at $\sim$1\,Hz and at variable frequency between 0.58 and 2.44\,Hz were found in two NS LMXBs, 4U\,1323--62 and EXO\,0748-676, respectively \citep{jonker1999, homan1999}. The PDS of these two sources (see Fig.\,2 in \citealt{jonker1999} and Fig.\,1 in \citealt{homan1999}) resemble the one of \source and the QPO central frequencies are consistent. However, 4U\,1323--62 and EXO\,0748-676 are dipping systems and the origin of the QPO is thought to be related to their high inclination. The light curves of \source did not show any dip and the inclination is unknown. From timing analysis that we performed, it is difficult to draw any conclusion about the origin of the QPO and the nature of the compact object of this new X-ray transient.
Even though at the beginning of the decay the RMS seems to be roughly constant while the hardness ratio clearly increases, when the source enters a fainter state associated with high hardness, the RMS drops. During the rebrightening a similar behaviour is detected: the RMS increases to 20$\%$ while  the hardness drops. This does not look like the typical behaviour of both accreting BH and NS, which show positive correlation between total RMS and hardness \citep{MunozDarias2011,MunozDarias2014}.

Another rough indicator on the nature of the source is its location on the radio:X-ray luminosity plane. BH and NS LMXBs in hard state are known to launch powerful jets, which can be observed in the radio-IR band. Plotting the BH transients and NS-LMXBs radio luminosity versus their X-ray luminosity, a clear correlation can be observed \citep[e.g.,][]{FenderKuulkers2001}. Additionally, BH and NS LMXBs populate two different regions (although with significant overlap) with the BH being systematically radio louder at the same X-ray luminosity \citep{VandenEijnden2021}. \source\ was observed by AMI-LA (central frequency of 15.5\,GHz) during the main outburst, on March 8 2023. No radio counterpart was detected and a radio flux density upper limit of $\sim$200 $\mu$Jy was established \citep{Atel15939}. The source was detected at a 0.5--10\,keV unabsorbed flux of $\sim$2.3$\times$10$^{-10}$\,\unitF on the same day by \nicer\, (ObsID 6203690102; Table \ref{tab:new_obs}). Since the distance of the source is unknown, we tentatively consider four values, that is 5, 10, 15 and 20\,kpc, and obtain four distinct results for the position of the source on the radio/X-rays luminosity plane (Fig. \ref{fig:lrlx}). Unfortunately, lacking any meaningful constraint on the source distance, the position of the source on the radio:X-ray luminosity diagram is rather ambiguous. If the source is located beyond 15\,kpc, the upper limit is consistent with \source being a distant BH, as also suggested in Section \ref{ss:disc-history} and by \citet{ATel15951}, although a radio bright NS can not be ruled out. This situation is reminiscent of what found for MAXI J1810--222, which was suggested to be a relatively distant BH transient, based also on the radio/X-rays location \citep{Russell2022}. If the system is instead closer, within 10 kpc, the source could be an exceptionally radio quiet BH X-ray binary or more likely a NS. 


\section{Conclusions}
In this study, we have analysed the 2023 outburst of the X-ray transient \source using data from \nicer, \nustar, and \swift. Our investigation focused on the spectral properties and temporal variability of the source to determine its nature and behaviour during the outburst. Our key findings can be summarised as follows:

$\bullet$ the source exhibited a complex outburst pattern with two major phases: an initial outburst followed by a rebrightening phase. During the outburst, the X-ray spectra were dominated by a Comptonisation component, roughly consistent with a power-law model with a photon index ranging from 1.5 to 2.2. The seed photons are likely provided by an accretion disc or a boundary layer, which is directly observed only during the brightest phases of the whole outburst, with temperature around 0.4\,keV. When the disc is not visible, the temperature of the seed photons is also poorly constrained, possibly becoming too cold to be detected by \nicer. An absorption feature around 0.96\,keV was consistently observed in both \nicer and \swift data, possibly indicating the presence of a wind, although its low significance prevents us to draw any solid interpretation on the nature of this feature;

$\bullet$ the broadband spectrum (0.5--60\,keV) was well-modelled by a combination of a multi-colour disc black-body and a Comptonised emission component, while we could not constrain the electronic temperature. The inclusion of a reflection component in the spectral model suggested a high ionisation parameter (${\rm log} X_i\sim 3.7$) and provided an upper limit for the inner radius of the accretion disc at $\sim$11.4 gravitational radii. The reflection fraction was not well constrained, indicating that reflection processes, although present, were not dominant;

$\bullet$ a QPO with a centroid frequency varying between $\sim$1.9 and 0.9\,Hz was detected and that anti-correlates with the hardness ratio was detected. The total RMS is constant while the hardness increases and the flux decreases, and then drops when the hardness is at its peak value;


$\bullet$ although no radio counterpart was detected, the radio luminosity upper limit, combined with X-ray luminosity data, positioned the source in a region on the radio:X-ray luminosity plane that is consistent both with a distant BH transient, and a radio bright NS.

Overall, our detailed analysis of \source during its 2023 outburst did not provide substantial evidence for the classification of the compact object. While the spectral analysis seems to point towards a BH nature, the timing results and the radio upper limit do not help us to rule out the NS scenario. 
Future more sensitive observations during the next outburst from this source, particularly in the optical and radio wavelengths, as well as continued X-ray monitoring, will be crucial to further constrain the nature of this intriguing X-ray transient and understand the mechanisms driving its complex outburst behaviour.

\begin{acknowledgements}
    The authors acknowledge financial contribution from the agreement ASI-INAF n.2017-14-H.0 and INAF mainstream (PI: A. De Rosa, T. Belloni), from the HERMES project financed by the Italian Space Agency (ASI) Agreement n. 2016/13 U.O and from the ASI-INAF Accordo Attuativo HERMES Technologic Pathfinder n. 2018-10-H.1-2020. We also acknowledge support from the European Union Horizon 2020 Research and Innovation Frame-work Programme under grant agreement HERMES-Scientific Pathfinder n. 821896 and from PRIN-INAF 2019 with the project "Probing the geometry of accretion: from theory to observations" (PI: Belloni).
 We thank the NuSTAR PI, Fiona Harrison, for approving the DDT request, and the NuSTAR SOC for carrying out the observation. We also thank Brad Cenko and the Swift duty scientists and science planners for making the Swift Target of Opportunity observations possible.
    A. Marino and F.C.Z. are supported by the H2020 ERC Consolidator Grant “MAGNESIA” under grant agreement No. 817661 (PI: Rea) and from grant SGR2021-01269 (PI: Graber/Rea). F.C.Z. is also supported by a Ram\'on y Cajal fellowship (grant agreement RYC2021-030888-I). 
    A.B. is supported by the Spanish Ministry of Science under the grant EUR2021-122010 (PI: Mu\~{n}oz-Darias), a L'Oreal--Unesco For Women In Science Fellowship (2023 Italian program) and ESA Fellowship.
    G.M. acknowledges financial support from the European Union's Horizon Europe research and innovation programme under the Marie Sk\l{}odowska-Curie grant agreement No. 101107057.
    M.A.P. acknowledges support through the Ram\'on y Cajal grant RYC2022-035388-I, funded by MCIU/AEI/10.13039/501100011033 and FSE+.
    T.M.-D. acknowledges support by the Spanish \textit{Agencia estatal de investigaci\'on} via PID2021-124879NB-I00.
    This work was also partially supported by the program Unidad de Excelencia Maria de Maeztu CEX2020-001058-M.
    M.C.B. acknowledges support from the INAF-Astrofit fellowship.
\end{acknowledgements}

\bibliographystyle{aa}
\bibliography{bibliography}

\onecolumn
\appendix

\section{Log of X-ray observations}
Table\,\ref{tab:new_obs} reports a journal of the X-ray observations of \source analysed in this work.


\begin{ThreePartTable}

\begin{TableNotes}
\footnotesize
\item [$*$] The instrumental setup is indicated in brackets: PC = photon counting, WT = windowed timing.
\item [$\dag$] Count rate in the 0.3--10\,keV range for \swift and \nicer, in the 3--60\,keV interval for \nustar. Uncertainties are at 1$\sigma$ c.l. 
\item [$\ddag$] These observations were merged for the spectral analysis.
\end{TableNotes}

\begin{longtable}{@{}cccccc}
\caption{\centering Log of the X-ray observations of \source analysed in this work.
\label{tab:new_obs}}\\

\hline\hline
X-ray Instrument$^*$ & Obs.ID & Start & Stop & Exposure &  Count Rate$^\dag$  \\
 & & \multicolumn{2}{c}{YYYY-MM-DD hh:mm:ss (TT)} & (ks) & (counts\,s$^{-1}$)  \\
 
\hline
\swift-XRT (PC) & 00015914001 & 2023-03-06 19:05:35 & 2023-03-06 21:50:29 & 3.4 & 0.48$\pm$0.01 \\

\nicer/XTI & 6203690102  & 2023-03-08 00:45:24 & 2023-03-08 22:48:40 & 6.1 & 26.54$\pm$0.08 \\
\nicer/XTI & 6203690104  & 2023-03-10 05:25:48 & 2023-03-10 21:15:41 & 7.6 & 25.98$\pm$0.08 \\

\nustar/FPMA & 90901309002  & 2023-03-10 05:41:09 & 2023-03-10 21:16:09 & 29.0 & 2.68$\pm$0.01 \\
\nustar/FPMB & 90901309002  & 2023-03-10 05:41:09 & 2023-03-10 21:16:09 & 28.8 & 2.494$\pm$0.009 \\

\swift-XRT (WT) & 00089604001 & 2023-03-10 10:06:00 & 2023-03-10 12:08:56 & 3.2 & 2.31$\pm$0.03 \\

\nicer/XTI & 6203690105  & 2023-03-11 03:07:49 & 2023-03-11 18:58:00 & 3.9 & 30.39$\pm$0.10 \\
\nicer/XTI & 6203690106  & 2023-03-12 08:34:21 & 2023-03-12 09:17:00 & 1.3 & 26.40$\pm$0.36 \\

\swift-XRT (WT) & 00089604002 & 2023-03-13 00:06:15 & 2023-03-13 06:47:55 & 2.9 & 1.34$\pm$0.02 \\

\nicer/XTI & 6203690110  & 2023-03-16 05:50:40 & 2023-03-16 20:11:40 & 2.5 & 24.75$\pm$0.11 \\

\swift-XRT (WT) & 00089604003 & 2023-03-16 13:50:09 & 2023-03-16 20:38:56 & 3.2 & 2.05$\pm$0.03 \\

\nicer/XTI & 6203690111  & 2023-03-17 00:06:14 & 2023-03-17 19:11:22 & 3.0 & 22.06$\pm$0.09 \\
\nicer/XTI & 6203690112  & 2023-03-18 15:17:00 & 2023-03-18 23:17:39 & 2.2 & 19.13$\pm$0.11 \\
\nicer/XTI & 6203690113  & 2023-03-19 03:33:00 & 2023-03-19 20:57:40 & 1.7 & 18.14$\pm$0.11 \\

\swift-XRT (WT) & 00089604004 & 2023-03-19 11:43:15 & 2023-03-19 21:39:56 & 3.1 & 1.13$\pm$0.02 \\

\nicer/XTI & 6203690114  & 2023-03-20 08:38:43 & 2023-03-20 18:38:40 & 3.3 & 17.29$\pm$0.08 \\

\swift-XRT (WT) & 00089604005 & 2023-03-22 06:33:04 & 2023-03-22 23:59:56 & 3.0 & 0.92$\pm$0.02 \\

\nicer/XTI & 6203690116  & 2023-03-22 10:14:33 & 2023-03-22 21:37:26 & 3.3 & 13.23$\pm$0.07 \\
\nicer/XTI & 6203690119  & 2023-03-25 18:49:50 & 2023-03-25 19:15:20 & 1.4 & 10.43$\pm$0.10 \\

\swift-XRT (WT) & 00089604006 & 2023-03-25T23:24:47 & 2023-03-25 23:50:56 & 1.7 & 0.73$\pm$0.02 \\

\nicer/XTI & 6203690120  & 2023-03-26 07:13:34 & 2023-03-26 19:58:40 & 1.6 & 9.43$\pm$0.08 \\
\nicer/XTI & 6203690122  & 2023-03-28 10:51:18 & 2023-03-28 15:21:30 & 1.0 & 8.60$\pm$0.10 \\

\swift-XRT (WT) & 00089604007 & 2023-03-29 16:09:28 & 2023-03-2916:36:56 & 1.6 & 0.60$\pm$0.02 \\
\swift-XRT (PC) & 00089604008 & 2023-03-31 03:11:06 & 2023-03-31 14:41:53 & 1.0 & 0.48$\pm$0.02 \\

\nicer/XTI & 6203690123  & 2023-03-31 22:03:41 & 2023-03-31 22:43:20 & 0.5 & 7.77$\pm$0.14 \\

\swift-XRT (PC) & 00089604009 & 2023-04-03 06:08:38 & 2023-04-03 17:26:53 & 1.3 & 0.39$\pm$0.02 \\

\nicer/XTI & 6203690124  & 2023-04-04 22:21:00 & 2023-04-04 22:50:40 & 1.4 & 5.63$\pm$0.07 \\

\swift-XRT (PC) & 00089604010 & 2023-04-06 02:09:44 & 2023-04-06 18:14:53 & 1.7 & 0.18$\pm$0.02 \\
\swift-XRT (PC) & 00089604011 & 2023-04-08 14:47:13 & 2023-04-08 20:59:53 & 2.9 & 0.32$\pm$0.01 \\

\nicer/XTI & 6203690130  & 2023-04-13 12:28:40 & 2023-04-13 14:24:20 & 1.8 & 3.39$\pm$0.05 \\

\swift-XRT (PC) & 00089604012 & 2023-04-15 07:12:23 & 2023-04-15 20:02:53 & 1.7 & 0.23$\pm$0.01 \\

\nicer/XTI & 6203690134  & 2023-04-18 15:59:39 & 2023-04-18 17:53:00 & 1.2 & 4.03$\pm$0.07 \\
\nicer/XTI & 6203690135  & 2023-04-20 13:04:00 & 2023-04-20 22:45:20 & 3.1 & 3.37$\pm$0.04 \\
\nicer/XTI & 6203690136  & 2023-04-21 07:32:40 & 2023-04-21 18:50:00 & 2.2 & 3.54$\pm$0.05 \\

\swift-XRT (PC) & 00089604013 & 2023-04-22 05:48:06 & 2023-04-22 13:59:53 & 2.7 & 0.23$\pm$0.01 \\

\nicer/XTI & 6203690137  & 2023-04-26 05:21:20 & 2023-04-26 14:59:20 & 3.1 & 3.28$\pm$0.04 \\
\nicer/XTI & 6203690139  & 2023-04-28 06:51:40 & 2023-04-28 23:51:45 & 0.8 & 4.19$\pm$0.08 \\

\swift-XRT (PC) & 00089604014 & 2023-04-28 15:29:00 & 2023-04-29 20:25:52 & 2.8 & 0.23$\pm$0.01 \\

\nicer/XTI & 6203690140  & 2023-04-29 13:50:20 & 2023-04-29 17:04:40 & 0.7 & 3.96$\pm$0.09 \\

\swift-XRT (PC) & 00089604015 & 2023-05-09 02:31:53 & 2023-05-09 23:03:53 & 2.4 & 0.19$\pm$0.01 \\

\nicer/XTI & 6203690145  & 2023-05-15 00:08:40 & 2023-05-15 14:30:40 & 1.9 & 2.68$\pm$0.05 \\

\swift-XRT (PC) & 00089604016 & 2023-05-16 06:09:22 & 2023-05-16 23:23:52 & 2.9 & 0.16$\pm$0.01 \\

\nicer/XTI & 6203690149  & 2023-05-19 06:18:08 & 2023-05-19 17:17:00 & 0.7 & 1.66$\pm$0.06 \\
\nicer/XTI & 6203690151  & 2023-05-21 01:39:15 & 2023-05-21 20:35:00 & 1.2 & 1.59$\pm$0.05 \\
\nicer/XTI & 6203690152  & 2023-05-22 10:24:20 & 2023-05-22 23:00:20 & 1.2 & 1.71$\pm$0.05 \\

\swift-XRT (PC) & 00089604017$^\ddag$ & 2023-05-24 02:57:29 & 2023-05-24 14:14:52 & 0.7 & 0.05$\pm$0.01 \\
\swift-XRT (PC) & 00089604018$^\ddag$ & 2023-05-25 00:56:26 & 2023-05-26 01:05:52 & 2.2 & 0.053$\pm$0.004 \\


\swift-XRT (PC) & 00089604019 & 2023-05-30 06:08:10 & 2023-05-30 20:33:52 & 3.2 & 0.037$\pm$0.004 \\
\swift-XRT (PC) & 00089604020 & 2023-06-01 03:04:52 & 2023-06-01 20:38:52 & 1.7 & 0.097$\pm$0.008 \\
\swift-XRT (PC) & 00089604021 & 2023-06-06 02:00:02 & 2023-06-06 22:34:53 & 2.2 & 0.21$\pm$0.01 \\


\swift-XRT (PC) & 00016073002 & 2023-06-11 02:35:40 & 2023-06-11 12:11:52 & 1.7 & 0.43$\pm$0.02 \\

\nicer/XTI & 6203690157  & 2023-06-14 09:13:00 & 2023-06-14 09:49:00 & 1.8 & 13.89$\pm$0.10 \\

\swift-XRT (PC) & 00016073003 & 2023-06-15 17:18:24 & 2023-06-15 17:42:53 & 1.5 & 0.43$\pm$0.02 \\

\nicer/XTI & 6203690159  & 2023-06-18 06:07:13 & 2023-06-18 06:36:51 & 1.1 & 16.65$\pm$0.13 \\

\swift-XRT (PC) & 00016073004 & 2023-06-19 00:52:31 & 2023-06-19 21:21:06 & 1.0 & 0.46$\pm$0.02 \\

\nicer/XTI & 6203690160  & 2023-06-19 05:38:00 & 2023-06-19 06:02:00 & 0.5 & 15.47$\pm$0.17 \\
\nicer/XTI & 6203690162  & 2023-06-22 01:48:40 & 2023-06-22 09:54:20 & 0.9 & 13.46$\pm$0.13 \\
\nicer/XTI & 6203690163  & 2023-06-23 02:14:21 & 2023-06-23 13:31:39 & 2.8 & 15.69$\pm$0.09 \\

\swift-XRT (PC) & 00016073005 & 2023-06-23 03:29:06 & 2023-06-23 12:55:56 & 0.9 & 0.46$\pm$0.02 \\

\nicer/XTI & 6203690165  & 2023-06-25 03:47:41 & 2023-06-25 16:30:40 & 0.7 & 18.48$\pm$0.18 \\

\swift-XRT (PC) & 00016073006 & 2023-06-27 02:20:55 & 2023-06-27 12:09:52 & 0.8 & 0.48$\pm$0.03 \\ 

\nicer/XTI & 6203690166  & 2023-06-27 08:34:00 & 2023-06-27 14:55:20 & 1.6 & 16.71$\pm$0.11 \\
\nicer/XTI & 6203690169  & 2023-07-02 04:42:30 & 2023-07-02 05:05:44 & 1.3 & 11.61$\pm$0.10 \\
\nicer/XTI & 6203690170  & 2023-07-03 10:08:25 & 2023-07-03 10:29:40 & 1.3 & 14.63$\pm$0.12 \\
\nicer/XTI & 6203690171  & 2023-07-04 04:57:01 & 2023-07-04 08:09:19 & 1.2 & 12.22$\pm$0.11 \\
\nicer/XTI & 6203690172  & 2023-07-06 23:31:36 & 2023-07-06 23:37:02 & 0.3 & 14.00$\pm$0.22 \\
\nicer/XTI & 6203690173  & 2023-07-09 08:45:07 & 2023-07-09 10:25:02 & 0.9 & 11.31$\pm$0.12 \\

\swift-XRT (PC) & 00016073007 & 2023-07-14 13:03:50 & 2023-07-15 22:57:53 & 0.4 & 0.68$\pm$0.04 \\

\nicer/XTI & 6203690179  & 2023-07-16 11:18:31 & 2023-07-16 11:22:20 & 0.2 & 11.08$\pm$0.23 \\
\nicer/XTI & 6203690180  & 2023-07-17 10:31:00 & 2023-07-17 10:38:00 & 0.4 & 11.94$\pm$0.18 \\
\nicer/XTI & 6203690181  & 2023-07-18 06:38:05 & 2023-07-18 06:39:04 & 0.06 & 12.32$\pm$0.48 \\

\swift-XRT (PC) & 00016073008 & 2023-07-19 07:34:28 & 2023-07-19 15:46:54 & 1.4 & 0.50$\pm$0.01 \\
\swift-XRT (PC) & 00016073009 & 2023-07-22 03:40:24 & 2023-07-22 15:01:52 & 0.5 & 0.48$\pm$0.03 \\

\nicer/XTI & 6203690185  & 2023-07-22 17:26:07 & 2023-07-22 20:46:00 & 1.4 & 11.23$\pm$0.10 \\
\nicer/XTI & 6203690186  & 2023-07-23 05:49:09 & 2023-07-23 15:21:00 & 1.7 & 6.45$\pm$0.07 \\
\nicer/XTI & 6203690187  & 2023-07-24 00:23:42 & 2023-07-24 08:23:00 & 1.2 & 4.78$\pm$0.07 \\

\swift-XRT (PC) & 00016073010$^\ddag$ & 2023-07-24 22:29:59 & 2023-07-24 22:32:52 & 0.2 & 0.56$\pm$0.06 \\     
\swift-XRT (PC) & 00016073011$^\ddag$ & 2023-07-25 01:48:54 & 2023-07-25 23:59:54 & 0.5 & 0.54$\pm$0.03 \\

\nicer/XTI & 6203690192  & 2023-07-29 12:10:17 & 2023-07-29 20:02:00 & 2.6 & 17.87$\pm$0.09 \\

\swift-XRT (PC) & 00016073012 & 2023-07-29 13:43:36 & 2023-07-29 13:46:52 & 0.2 & 0.54$\pm$0.05 \\

\nicer/XTI & 6203690193  & 2023-07-30 09:39:53 & 2023-07-30 14:22:48 & 1.0 & 18.31$\pm$0.14 \\
\nicer/XTI & 6203690195  & 2023-08-01 00:15:48 & 2023-08-01 03:32:01 & 0.9 & 20.55$\pm$0.16 \\
\nicer/XTI & 6203690196  & 2023-08-02 21:16:52 & 2023-08-02 21:20:35 & 0.2 & 20.03$\pm$0.31 \\

\swift-XRT (PC) & 00016073013 & 2023-08-04 04:09:00 & 2023-08-04 21:59:52 & 1.8 & 0.52$\pm$0.02 \\
\swift-XRT (PC) & 00016073014 & 2023-08-11 02:29:36 & 2023-08-11 18:36:53 & 1.6 & 0.53$\pm$0.02 \\

\nicer/XTI & 6203690203  & 2023-08-17 09:53:01 & 2023-08-17 11:39:20 & 1.0 & 11.83$\pm$0.12 \\

\swift-XRT (PC) & 00016073015 & 2023-08-18 02:58:20 & 2023-08-19 23:24:52 & 1.4 & 0.52$\pm$0.02 \\

\nicer/XTI & 6203690204  & 2023-08-18 12:16:27 & 2023-08-18 15:28:20 & 0.7 & 12.99$\pm$0.20 \\
\nicer/XTI & 6203690205  & 2023-08-19 06:49:26 & 2023-08-19 16:24:10 & 2.0 & 13.13$\pm$0.13 \\
\nicer/XTI & 6203690207  & 2023-08-21 03:51:13 & 2023-08-21 03:54:21 & 0.2 & 12.48$\pm$0.27 \\

\swift-XRT (PC) & 00016073016 & 2023-08-25 02:56:06 & 2023-08-25 09:17:53 & 1.7 & 0.58$\pm$0.02 \\

\nicer/XTI & 6203690209  & 2023-08-26 23:22:35 & 2023-08-26 23:33:00 & 0.6 & 11.56$\pm$0.14 \\
\nicer/XTI & 6203690210  & 2023-08-27 13:20:24 & 2023-08-27 13:29:20 & 0.5 & 11.32$\pm$0.15 \\
\nicer/XTI & 6203690213  & 2023-08-30 09:33:32 & 2023-08-30 09:38:40 & 0.3 & 9.74$\pm$0.19 \\

\swift-XRT (PC) & 00016073017 & 2023-09-01 07:33:29 & 2023-09-01 07:47:52 & 0.9 & 0.51$\pm$0.02 \\

\nicer/XTI & 6203690215  & 2023-09-01 11:08:55 & 2023-09-01 23:37:20 & 2.5 & 11.83$\pm$0.08 \\
\nicer/XTI & 6203690219  & 2023-09-05 12:38:50 & 2023-09-05 12:45:20 & 0.4 & 11.23$\pm$0.18 \\
\nicer/XTI & 6203690225  & 2023-09-11 23:24:02 & 2023-09-11 23:32:40 & 0.5 & 9.26$\pm$0.14 \\

\swift-XRT (PC) & 00016073018 & 2023-09-12 10:07:32 & 2023-09-12 13:16:52 & 1.0 & 0.44$\pm$0.02 \\

\nicer/XTI & 6203690226  & 2023-09-12 18:03:08 & 2023-09-12 18:13:26 & 0.4 & 11.05$\pm$0.19 \\

\swift-XRT (PC) & 00016073019 & 2023-09-18 00:47:05 & 2023-09-18 12:11:52 & 2.1 & 0.46$\pm$0.01 \\

\nicer/XTI & 6203690227  & 2023-09-21 15:49:34 & 2023-09-21 15:53:20 & 0.2 & 7.57$\pm$0.20 \\
\nicer/XTI & 6203690228  & 2023-09-24 13:39:36 & 2023-09-24 16:48:53 & 0.7 & 6.01$\pm$0.10 \\

\swift-XRT (PC) & 00016073020 & 2023-09-25 20:04:17 & 2023-09-25 23:27:53 & 1.7 & 0.36$\pm$0.01 \\

\nicer/XTI & 6203690229  & 2023-09-29 00:22:34 & 2023-09-29 20:39:40 & 1.5 & 8.27$\pm$0.08 \\

\swift-XRT (PC) & 00016073021 & 2023-10-02 13:26:23 & 2023-10-02 20:10:53 & 2.1 & 0.46$\pm$0.01 \\

\hline


\insertTableNotes  

\end{longtable}

\end{ThreePartTable}



\section{Spectral analysis results}

Tables\,\ref{tab:single-nicer-1a} and \ref{tab:single-nicer-1b} report the results of the spectral analysis of the \nicer spectra, while we listed the best-fitting parameters for the \swift-XRT dataset in Table\,\ref{tab:spec_swift}.

\begin{longtable}{@{}ccccccccc}
\caption{Best fit values derived from the modelling of the single \nicer spectra using Model 0 or Model 1A, according to the statistical significance of the fit in each case. $K_{\rm nthc}$ and $K_{\rm bb}$ are the normalisations of the \texttt{nthcomp} and \texttt{bbodyrad} components, respectively. The unabsorbed flux is estimated in the 0.5--10\,keV energy interval. 
\label{tab:single-nicer-1a}} \\
\hline
\hline
ObsID & $\Gamma$ & $K_{\rm nthc}\times$100 & $kT_{\rm seed}$ & $kT_{\rm bbody}$ & $K_{\rm bb}$ & $F_{0510}$ & $\chi^2$ \T \B \\ 
      &          &                     & (keV)       &   (keV) &       & ($\times$10$^{-10}$ erg cm$^{-2}$ s$^{-1}$) & (dof) \T \B \\ 
\hline
6203690102 & 2.01$\pm$0.04 &3.887$\pm$0.297 &0.22$^{+0.01}_{-0.01}$ &=kT$_{\rm seed}$ & 1827$^{+255}_{-200}$ &2.34$\pm$0.02 &137(96)\T \B \\  
6203690104 & 2.23$\pm$0.05 &4.560$^{+0.680}_{-0.435}$ &0.19$^{+0.02}_{-0.04}$ &=kT$_{\rm seed}$ & 1188$^{+268}_{-1188}$ &2.04$\pm$0.02 &74(72)\T \B \\  
6203690105 & 1.95$\pm$0.04 &3.280$\pm$0.253 &0.20$^{+0.01}_{-0.01}$ &=kT$_{\rm seed}$ & 1768$^{+385}_{-266}$ &1.98$\pm$0.01 &133(98)\T \B \\  
6203690106 & 2.17$^{+0.04}_{-0.03}$ &4.430$\pm$0.082 &$<$0.23&- & - & 1.88$\pm$0.02 &55(66)\T \B \\  
6203690110 & 1.84$\pm$0.05 &2.380$\pm$0.220 &0.21$^{+0.01}_{-0.01}$ &=kT$_{\rm seed}$ & 1362$^{+300}_{-215}$ &1.67$\pm$0.02 &90(93)\T \B \\  
6203690111 & 1.85$\pm$0.04 &2.265$\pm$0.203 &0.19$^{+0.02}_{-0.02}$ &=kT$_{\rm seed}$ & 1401$^{+526}_{-298}$ &1.46$\pm$0.01 &116(96)\T \B \\  
6203690112 & 1.86$\pm$0.06 &2.012$\pm$0.248 &0.19$^{+0.02}_{-0.03}$ &=kT$_{\rm seed}$ & 1171$^{+804}_{-330}$ &1.26$\pm$0.02 &97(83)\T \B \\  
6203690113 & 1.93$\pm$0.02 &2.397$\pm$0.033 &$<$0.11&- & - & 1.22$\pm$0.01 &100(90)\T \B \\  
6203690114 & 1.77$^{+0.04}_{-0.05}$ &1.663$\pm$0.154 &0.20$^{+0.02}_{-0.02}$ &=kT$_{\rm seed}$ & 866$^{+301}_{-179}$ &1.18$\pm$0.01 &110(97)\T \B \\  
6203690116 & 1.66$\pm$0.05 &1.183$\pm$0.117 &0.20$^{+0.02}_{-0.02}$ &=kT$_{\rm seed}$ & 618$^{+312}_{-160}$ &0.93$\pm$0.01 &108(96)\T \B \\  
6203690119 & 1.66$\pm$0.04 &1.023$\pm$0.031 &$<$0.12&- & - & 0.68$\pm$0.02 &120(80)\T \B \\  
6203690120 & 1.70$\pm$0.03 &0.994$\pm$0.013 &$<$0.13&- & - & 0.64$\pm$0.01 &110(89)\T \B \\  
6203690122 & 1.71$\pm$0.04 &0.920$\pm$0.031 &$<$0.13&- & - & 0.58$\pm$0.01 &88(78)\T \B \\  
6203690123 & 1.66$^{+0.09}_{-0.07}$ &0.765$\pm$0.027 &$<$0.24&- & - & 0.51$\pm$0.02 &61(56)\T \B \\  
6203690124 & 1.45$^{+0.05}_{-0.04}$ &0.517$\pm$0.021 &$<$0.15&- & - & 0.44$\pm$0.01 &79(83)\T \B \\  
6203690130 & 1.39$^{+0.09}_{-0.07}$ &0.269$\pm$0.027 &$<$0.40&- & - & 0.26$\pm$0.01 &57(74)\T \B \\  
6203690131 & 1.49$^{+0.07}_{-0.06}$ &0.307$\pm$0.029 &0.22$^{+0.06}_{-0.22}$ &- & - & 0.27$\pm$0.01 &77(84)\T \B \\  
6203690134 & 1.58$^{+0.12}_{-0.09}$ &0.368$\pm$0.030 &$<$0.37&- & - & 0.28$\pm$0.01 &56(59)\T \B \\  
6203690135 & 1.34$^{+0.04}_{-0.03}$ &0.272$\pm$0.009 &$<$0.19&- & - & 0.27$\pm$0.01 &78(88)\T \B \\  
6203690136 & 1.40$^{+0.06}_{-0.05}$ &0.286$\pm$0.021 &$<$0.36&- & - & 0.27$\pm$0.01 &70(85)\T \B \\  
6203690137 & 1.40$\pm$0.04 &0.279$\pm$0.010 &$<$0.19&- & - & 0.26$\pm$0.01 &58(80)\T \B \\  
6203690138 & 1.32$^{+0.06}_{-0.07}$ &0.272$\pm$0.015 &$<$0.18&- & - & 0.28$\pm$0.01 &49(62)\T \B \\  
6203690139 & 1.28$^{+0.07}_{-0.06}$ &0.306$\pm$0.020 &$<$0.22&- & - & 0.33$\pm$0.02 &73(71)\T \B \\  
6203690140 & 1.39$^{+0.09}_{-0.08}$ &0.313$\pm$0.020 &$<$0.17&- & - & 0.29$\pm$0.02 &63(57)\T \B \\  
6203690145 & 1.33$^{+0.06}_{-0.05}$ &0.217$\pm$0.014 &$<$0.22&- & - & 0.22$\pm$0.01 &65(73)\T \B \\  
6203690149 & 1.87$^{+0.74}_{-0.35}$ &0.117$\pm$0.030 &0.36$^{+0.15}_{-0.18}$ &- & - & 0.10$\pm$0.01 &52(52)\T \B \\  
6203690151 & 1.00$^{+0.17}_{-1.00}$ &0.063$\pm$0.021 &$<$0.74&- & - & 0.14$\pm$0.01 &57(58)\T \B \\  
6203690152 & 1.00$^{+0.07}_{-1.00}$ &0.093$\pm$0.007 &$<$0.27&- & - & 0.16$\pm$0.01 &48(57)\T \B \\  
6203690157 & 1.96$^{+0.03}_{-0.02}$ &1.748$\pm$0.028 &$<$0.11&- & - & 0.87$\pm$0.01 &119(90)\T \B \\  
6203690159 & 1.78$\pm$0.09 &1.354$\pm$0.220 &0.23$^{+0.02}_{-0.02}$ &=kT$_{\rm seed}$ & 654$^{+243}_{-145}$ &1.03$\pm$0.02 &96(88)\T \B \\  
6203690160 & 1.95$^{+0.05}_{-0.04}$ &1.951$\pm$0.061 &$<$0.22&- & - & 0.97$\pm$0.02 &111(79)\T \B \\  
6203690162 & 1.80$^{+0.09}_{-0.10}$ &1.382$\pm$0.247 &0.21$^{+0.02}_{-0.03}$ &=kT$_{\rm seed}$ & 745$^{+443}_{-216}$ &0.99$\pm$0.02 &83(81)\T \B \\  
6203690163 & 1.76$\pm$0.06 &1.358$\pm$0.140 &0.21$^{+0.02}_{-0.02}$ &=kT$_{\rm seed}$ & 877$^{+287}_{-181}$ &1.01$\pm$0.01 &93(87)\T \B \\  
6203690165 & 1.69$^{+0.09}_{-0.10}$ &1.517$\pm$0.224 &0.20$^{+0.02}_{-0.02}$ &=kT$_{\rm seed}$ & 1484$^{+1013}_{-485}$ &1.21$\pm$0.03 &55(79)\T \B \\  
6203690166 & 1.79$\pm$0.07 &1.417$\pm$0.195 &0.22$^{+0.02}_{-0.02}$ &=kT$_{\rm seed}$ & 793$^{+311}_{-178}$ &1.04$\pm$0.01 &110(88)\T \B \\  
6203690169 & 1.77$^{+0.08}_{-0.09}$ &1.176$\pm$0.199 &0.21$^{+0.02}_{-0.03}$ &=kT$_{\rm seed}$ & 598$^{+327}_{-163}$ &0.86$\pm$0.02 &88(87)\T \B \\  
6203690170 & 1.81$\pm$0.08 &1.263$\pm$0.183 &0.21$^{+0.02}_{-0.02}$ &=kT$_{\rm seed}$ & 788$^{+363}_{-198}$ &0.91$\pm$0.02 &120(88)\T \B \\  
6203690171 & 1.93$\pm$0.03 &1.591$\pm$0.040 &$<$0.13&- & - & 0.81$\pm$0.01 &92(88)\T \B \\  
6203690172 & 1.87$^{+0.06}_{-0.05}$ &1.627$\pm$0.072 &$<$0.15&- & - & 0.88$\pm$0.03 &95(77)\T \B \\  
6203690173 & 1.89$\pm$0.04 &1.359$\pm$0.173 &$<$0.13&- & - & 0.72$\pm$0.01 &86(86)\T \B \\  
6203690179 & 1.97$^{+0.12}_{-0.10}$ &1.299$\pm$0.098 &$<$0.34&- & - & 0.65$\pm$0.03 &102(73)\T \B \\  
6203690180 & 1.18$^{+0.10}_{-0.12}$ &0.716$\pm$0.114 &0.17$^{+0.04}_{-0.04}$ &=kT$_{\rm seed}$ & 1814$^{+6148}_{-1217}$ &1.02$\pm$0.04 &106(80)\T \B \\  
6203690181 & 1.99$^{+0.43}_{-0.27}$ &1.235$\pm$0.234 &$<$0.44&- & - & 0.66$\pm$0.07 &62(47)\T \B \\  
6203690185 & 1.80$\pm$0.03 &1.270$\pm$0.030 &$<$0.14&- & - & 0.73$\pm$0.01 &110(88)\T \B \\  
6203690186 & 1.72$\pm$0.09 &1.006$\pm$0.165 &0.21$^{+0.03}_{-0.03}$ &=kT$_{\rm seed}$ & 565$^{+370}_{-176}$ &0.77$\pm$0.02 &77(86)\T \B \\  
6203690187 & 1.83$\pm$0.05 &1.324$\pm$0.048 &$<$0.16&- & - & 0.74$\pm$0.02 &73(79)\T \B \\  
6203690192 & 1.79$\pm$0.05 &1.548$\pm$0.149 &0.21$^{+0.01}_{-0.02}$ &=kT$_{\rm seed}$ & 1035$^{+317}_{-204}$ &1.12$\pm$0.01 &108(95)\T \B \\  
6203690193 & 1.81$\pm$0.07 &1.672$\pm$0.210 &0.20$^{+0.02}_{-0.02}$ &=kT$_{\rm seed}$ & 1187$^{+655}_{-334}$ &1.15$\pm$0.02 &71(88)\T \B \\  
6203690195 & 1.89$^{+0.08}_{-0.09}$ &1.775$\pm$0.262 &0.22$^{+0.02}_{-0.02}$ &=kT$_{\rm seed}$ & 1119$^{+363}_{-232}$ &1.23$\pm$0.02 &92(87)\T \B \\  
6203690196 & 1.64$^{+0.14}_{-0.16}$ &1.590$\pm$0.387 &0.19$^{+0.04}_{-0.04}$ &=kT$_{\rm seed}$ & 1693$^{+3531}_{-811}$ &1.33$\pm$0.05 &72(77)\T \B \\  
6203690203 & 1.81$^{+0.10}_{-0.12}$ &1.154$\pm$0.249 &0.22$^{+0.03}_{-0.04}$ &=kT$_{\rm seed}$ & 554$^{+400}_{-167}$ &0.82$\pm$0.02 &61(78)\T \B \\  
6203690204 & 2.08$^{+0.08}_{-0.06}$ &1.768$\pm$0.052 &$<$0.18&- & - & 0.79$\pm$0.02 &45(65)\T \B \\  
6203690205 & 1.42$^{+0.15}_{-0.16}$ &0.734$\pm$0.160 &0.24$^{+0.02}_{-0.02}$ &=kT$_{\rm seed}$ & 511$^{+175}_{-112}$ &0.89$\pm$0.02 &36(68)\T \B \\  
6203690207 & 2.09$\pm$0.08 &1.659$\pm$0.093 &$<$0.14&- & - & 0.74$\pm$0.03 &72(72)\T \B \\  
6203690209 & 2.05$\pm$0.04 &1.530$\pm$0.050 &$<$0.14&- & - & 0.71$\pm$0.02 &88(81)\T \B \\  
6203690210 & 1.97$^{+0.06}_{-0.05}$ &1.434$\pm$0.044 &$<$0.20&- & - & 0.70$\pm$0.02 &55(79)\T \B \\  
6203690213 & 2.01$\pm$0.07 &1.238$\pm$0.059 &$<$0.16&- & - & 0.59$\pm$0.02 &90(74)\T \B \\  
6203690215 & 1.72$^{+0.06}_{-0.07}$ &0.982$\pm$0.120 &0.21$^{+0.02}_{-0.02}$ &=kT$_{\rm seed}$ & 581$^{+255}_{-142}$ &0.76$\pm$0.01 &107(90)\T \B \\  
6203690219 & 2.04$\pm$0.06 &1.446$\pm$0.067 &$<$0.23&- & - & 0.67$\pm$0.02 &89(77)\T \B \\  
6203690225 & 2.01$\pm$0.06 &1.186$\pm$0.045 &$<$0.14&- & - & 0.57$\pm$0.02 &84(77)\T \B \\  
6203690226 & 1.68$\pm$0.07 &1.086$\pm$0.057 &$<$0.13&- & - & 0.71$\pm$0.03 &144(79)\T \B \\  
6203690227 & 1.73$^{+0.10}_{-0.09}$ &0.869$\pm$0.064 &$<$0.16&- & - & 0.54$\pm$0.03 &52(72)\T \B \\  
6203690228 & 1.71$\pm$0.06 &0.627$\pm$0.030 &$<$0.15&- & - & 0.40$\pm$0.01 &89(78)\T \B \\  
6203690229 & 1.92$^{+0.04}_{-0.03}$ &1.015$\pm$0.028 &$<$0.13&- & - & 0.52$\pm$0.01 &77(88)\T \B \\  
\hline
\hline
\end{longtable}
 
\begin{longtable}{@{}ccccccccc}
\caption{Best fit values derived from the modelling of the single \nicer spectra using Model 0 or Model 1B, according to the statistical significance of the fit in each case. $K_{\rm nthc}$ and $K_{\rm disc}$ are the normalisations of the \texttt{nthcomp} and \texttt{diskbb} components, respectively. The unabsorbed flux is estimated in the 0.5--10\,keV energy interval. 
\label{tab:single-nicer-1b}} \\
\hline
\hline
ObsID & $\Gamma$ & $K_{\rm nthc}\times$100 & $kT_{\rm seed}$ & $kT_{\rm disk}$ & $K_{\rm disk}$ & $F_{0510}$ & $\chi^2$ \T \B \\ 
      &          &                     & (keV)       &   (keV) &       & ($\times$10$^{-10}$ erg cm$^{-2}$ s$^{-1}$) & (dof) \T \B \\ 
\hline
6203690102 & 1.93$\pm$0.05 &3.501$\pm$0.310 &0.34$^{+0.02}_{-0.02}$ &=$kT_{\rm seed}$ & 258$^{+46}_{-35}$ &2.41$\pm$0.02 &109(96)\T \B \\  
6203690104 & 2.19$^{+0.07}_{-0.08}$ &4.153$\pm$0.564 &0.31$^{+0.03}_{-0.04}$ &=$kT_{\rm seed}$ & 188$^{+40}_{-39}$ &2.08$\pm$0.02 &72(72)\T \B \\  
6203690105 & 1.88$\pm$0.05 &3.016$\pm$0.270 &0.31$^{+0.02}_{-0.02}$ &=$kT_{\rm seed}$ & 253$^{+68}_{-47}$ &2.02$\pm$0.01 &114(98)\T \B \\  
6203690106 & 2.09$^{+0.09}_{-0.12}$ &3.565$\pm$0.812 &0.27$^{+0.05}_{-0.10}$ &=$kT_{\rm seed}$ & 307$^{+324}_{-72}$ &1.84$\pm$0.03 &52(65)\T \B \\  
6203690110 & 1.77$\pm$0.07 &2.204$\pm$0.247 &0.33$^{+0.02}_{-0.03}$ &=$kT_{\rm seed}$ & 202$^{+61}_{-42}$ &1.72$\pm$0.02 &79(93)\T \B \\  
6203690111 & 1.80$\pm$0.06 &2.115$\pm$0.217 &0.30$^{+0.03}_{-0.03}$ &=$kT_{\rm seed}$ & 201$^{+92}_{-52}$ &1.49$\pm$0.01 &104(96)\T \B \\  
6203690112 & 1.80$^{+0.08}_{-0.09}$ &1.852$\pm$0.271 &0.30$^{+0.04}_{-0.05}$ &=$kT_{\rm seed}$ & 159$^{+121}_{-52}$ &1.28$\pm$0.02 &92(83)\T \B \\  
6203690113 & 1.82$^{+0.08}_{-0.09}$ &1.893$\pm$0.328 &0.28$^{+0.05}_{-0.08}$ &=$kT_{\rm seed}$ & 163$^{+403}_{-68}$ &1.24$\pm$0.02 &92(89)\T \B \\  
6203690114 & 1.71$\pm$0.06 &1.563$\pm$0.168 &0.32$^{+0.03}_{-0.03}$ &=$kT_{\rm seed}$ & 119$^{+54}_{-31}$ &1.21$\pm$0.01 &100(97)\T \B \\  
6203690116 & 1.62$\pm$0.06 &1.159$\pm$0.125 &0.30$^{+0.04}_{-0.04}$ &=$kT_{\rm seed}$ & 94$^{+70}_{-32}$ &0.95$\pm$0.01 &103(96)\T \B \\  
6203690119 & 1.51$^{+0.09}_{-0.14}$ &0.826$\pm$0.119 &0.25$^{+0.10}_{-0.08}$ &=$kT_{\rm seed}$ & 148$^{+1018}_{-106}$ &0.73$\pm$0.02 &102(79)\T \B \\  
6203690120 & 1.70$\pm$0.03 &0.994$\pm$0.026 &$<$0.18&- & - & 0.64$\pm$0.01 &110(89)\T \B \\  
6203690122 & 1.71$\pm$0.04 &0.921$\pm$0.028 &$<$0.20&- & - & 0.58$\pm$0.01 &88(78)\T \B \\  
6203690123 & 1.65$^{+0.10}_{-0.07}$ &0.765$\pm$0.036 &$<$0.34&- & - & 0.51$\pm$0.02 &61(56)\T \B \\  
6203690124 & 1.26$^{+0.14}_{-0.20}$ &0.375$\pm$0.104 &0.34$^{+0.15}_{-0.17}$ &=$kT_{\rm seed}$ & 21$^{+456}_{-15}$ &0.47$\pm$0.02 &67(82)\T \B \\  
6203690130 & 1.41$^{+0.14}_{-0.08}$ &0.269$\pm$0.024 &$<$0.80&- & - & 0.26$\pm$0.01 &57(74)\T \B \\  
6203690131 & 1.52$^{+0.12}_{-0.08}$ &0.307$\pm$0.022 &0.43$^{+0.18}_{-0.22}$ &- & - & 0.27$\pm$0.01 &75(84)\T \B \\  
6203690134 & 1.64$^{+0.24}_{-0.13}$ &0.359$\pm$0.037 &$<$0.79&- & - & 0.27$\pm$0.01 &55(59)\T \B \\  
6203690135 & 1.34$\pm$0.04 &0.272$\pm$0.007 &$<$0.38&- & - & 0.27$\pm$0.01 &78(88)\T \B \\  
6203690136 & 1.41$^{+0.08}_{-0.06}$ &0.288$\pm$0.009 &$<$0.60&- & - & 0.27$\pm$0.01 &70(85)\T \B \\  
6203690137 & 1.40$\pm$0.04 &0.279$\pm$0.009 &$<$0.34&- & - & 0.26$\pm$0.01 &58(80)\T \B \\  
6203690138 & 1.32$\pm$0.06 &0.272$\pm$0.015 &$<$0.35&- & - & 0.28$\pm$0.01 &49(62)\T \B \\  
6203690139 & 1.29$\pm$0.07 &0.306$\pm$0.019 &$<$0.47&- & - & 0.33$\pm$0.02 &73(71)\T \B \\  
6203690140 & 1.39$^{+0.09}_{-0.08}$ &0.312$\pm$0.022 &$<$0.31&- & - & 0.29$\pm$0.02 &63(57)\T \B \\  
6203690145 & 1.33$^{+0.06}_{-0.05}$ &0.217$\pm$0.011 &$<$0.44&- & - & 0.22$\pm$0.01 &65(73)\T \B \\  
6203690149 & $<$1.70 &0.121$\pm$0.017 &$<$2.20&- & - & 0.09$\pm$0.01 &48(52)\T \B \\  
6203690151 & $<$1.03 &0.067$\pm$0.009 &$<$2.41&- & - & 0.14$\pm$0.01 &54(58)\T \B \\  
6203690152 & $<$1.07 &0.093$\pm$0.007 &$<$1.10&- & - & 0.16$\pm$0.01 &48(57)\T \B \\  
6203690157 & 1.96$\pm$0.02 &1.750$\pm$0.035 &$<$0.15&- & - & 0.87$\pm$0.01 &119(90)\T \B \\  
6203690159 & 1.71$^{+0.11}_{-0.13}$ &1.282$\pm$0.251 &0.35$^{+0.04}_{-0.05}$ &=$kT_{\rm seed}$ & 97$^{+54}_{-29}$ &1.06$\pm$0.02 &93(88)\T \B \\  
6203690160 & 1.95$^{+0.05}_{-0.04}$ &1.951$\pm$0.062 &$<$0.28&- & - & 0.97$\pm$0.02 &111(79)\T \B \\  
6203690162 & 1.75$^{+0.12}_{-0.14}$ &1.333$\pm$0.284 &0.32$^{+0.05}_{-0.06}$ &=$kT_{\rm seed}$ & 114$^{+110}_{-44}$ &1.01$\pm$0.02 &83(81)\T \B \\  
6203690163 & 1.67$^{+0.08}_{-0.09}$ &1.234$\pm$0.165 &0.32$^{+0.03}_{-0.03}$ &=$kT_{\rm seed}$ & 125$^{+56}_{-33}$ &1.04$\pm$0.01 &84(87)\T \B \\  
6203690165 & 1.64$^{+0.11}_{-0.13}$ &1.475$\pm$0.260 &0.28$^{+0.05}_{-0.05}$ &=$kT_{\rm seed}$ & 284$^{+327}_{-126}$ &1.24$\pm$0.03 &55(79)\T \B \\  
6203690166 & 1.70$^{+0.10}_{-0.11}$ &1.270$\pm$0.215 &0.34$^{+0.04}_{-0.04}$ &=$kT_{\rm seed}$ & 109$^{+55}_{-31}$ &1.07$\pm$0.02 &101(88)\T \B \\  
6203690169 & 1.71$^{+0.11}_{-0.13}$ &1.098$\pm$0.221 &0.33$^{+0.05}_{-0.06}$ &=$kT_{\rm seed}$ & 84$^{+65}_{-29}$ &0.88$\pm$0.02 &85(87)\T \B \\  
6203690170 & 1.76$^{+0.09}_{-0.10}$ &1.221$\pm$0.200 &0.31$^{+0.04}_{-0.04}$ &=$kT_{\rm seed}$ & 131$^{+89}_{-44}$ &0.93$\pm$0.02 &117(88)\T \B \\  
6203690171 & 1.93$\pm$0.03 &1.592$\pm$0.040 &0.01$^{+0.18}_{-0.01}$ &- & - & 0.81$\pm$0.01 &92(88)\T \B \\  
6203690172 & 1.87$^{+0.06}_{-0.05}$ &1.626$\pm$0.053 &$<$0.22&- & - & 0.88$\pm$0.03 &95(77)\T \B \\  
6203690173 & 1.89$\pm$0.04 &1.360$\pm$0.041 &$<$0.19&- & - & 0.72$\pm$0.01 &86(86)\T \B \\  
6203690179 & 1.97$^{+0.14}_{-0.10}$ &1.307$\pm$0.094 &$<$0.50&- & - & 0.66$\pm$0.03 &102(73)\T \B \\  
6203690180 & 1.11$^{+0.14}_{-1.11}$ &0.656$\pm$0.145 &0.26$^{+0.08}_{-0.08}$ &=$kT_{\rm seed}$ & 257$^{+1273}_{-210}$ &1.02$\pm$0.04 &104(80)\T \B \\  
6203690181 & 2.06$^{+1.16}_{-0.32}$ &1.240$\pm$0.230 &$<$0.79&- & - & 0.66$\pm$0.07 &61(47)\T \B \\  
6203690185 & 1.80$\pm$0.03 &1.270$\pm$0.030 &$<$0.20&- & - & 0.73$\pm$0.01 &110(88)\T \B \\  
6203690186 & 1.68$^{+0.11}_{-0.13}$ &0.996$\pm$0.190 &0.31$^{+0.06}_{-0.06}$ &=$kT_{\rm seed}$ & 92$^{+108}_{-40}$ &0.79$\pm$0.02 &78(86)\T \B \\  
6203690187 & 1.83$\pm$0.05 &1.324$\pm$0.129 &$<$0.25&- & - & 0.74$\pm$0.02 &73(79)\T \B \\  
6203690192 & 1.72$\pm$0.07 &1.438$\pm$0.165 &0.32$^{+0.03}_{-0.03}$ &=$kT_{\rm seed}$ & 157$^{+64}_{-39}$ &1.15$\pm$0.01 &97(95)\T \B \\  
6203690193 & 1.77$^{+0.08}_{-0.10}$ &1.632$\pm$0.244 &0.29$^{+0.04}_{-0.05}$ &=$kT_{\rm seed}$ & 205$^{+187}_{-79}$ &1.17$\pm$0.02 &70(88)\T \B \\  
6203690195 & 1.82$^{+0.11}_{-0.12}$ &1.660$\pm$0.306 &0.33$^{+0.04}_{-0.04}$ &=$kT_{\rm seed}$ & 178$^{+87}_{-50}$ &1.27$\pm$0.02 &90(87)\T \B \\  
6203690196 & 1.57$^{+0.18}_{-0.23}$ &1.481$\pm$0.453 &0.29$^{+0.08}_{-0.09}$ &=$kT_{\rm seed}$ & 279$^{+1016}_{-170}$ &1.36$\pm$0.05 &71(77)\T \B \\  
6203690203 & 1.76$^{+0.13}_{-0.16}$ &1.120$\pm$0.272 &0.32$^{+0.06}_{-0.07}$ &=$kT_{\rm seed}$ & 87$^{+98}_{-35}$ &0.84$\pm$0.02 &59(78)\T \B \\  
6203690204 & 2.08$^{+0.08}_{-0.06}$ &1.767$\pm$0.062 &$<$0.24&- & - & 0.79$\pm$0.02 &45(65)\T \B \\  
6203690207 & 2.08$^{+0.08}_{-0.07}$ &1.655$\pm$0.094 &$<$0.20&- & - & 0.74$\pm$0.03 &72(72)\T \B \\  
6203690209 & 2.05$^{+0.05}_{-0.04}$ &1.530$\pm$0.059 &$<$0.20&- & - & 0.71$\pm$0.02 &88(81)\T \B \\  
6203690210 & 1.97$^{+0.06}_{-0.05}$ &1.434$\pm$0.050 &$<$0.25&- & - & 0.70$\pm$0.02 &56(79)\T \B \\  
6203690213 & 2.01$\pm$0.07 &1.238$\pm$0.040 &$<$0.23&- & - & 0.59$\pm$0.02 &90(74)\T \B \\  
6203690215 & 1.64$^{+0.09}_{-0.10}$ &0.900$\pm$0.138 &0.33$^{+0.04}_{-0.04}$ &=$kT_{\rm seed}$ & 80$^{+49}_{-25}$ &0.78$\pm$0.01 &100(90)\T \B \\  
6203690219 & 2.04$^{+0.09}_{-0.07}$ &1.441$\pm$0.071 &$<$0.35&- & - & 0.67$\pm$0.02 &89(77)\T \B \\  
6203690225 & 2.00$^{+0.06}_{-0.05}$ &1.185$\pm$0.052 &$<$0.20&- & - & 0.57$\pm$0.02 &84(77)\T \B \\  
6203690226 & 1.25$^{+0.23}_{-1.25}$ &0.605$\pm$0.288 &0.31$^{+0.12}_{-0.11}$ &=$kT_{\rm seed}$ & 107$^{+681}_{-78}$ &0.83$\pm$0.04 &118(78)\T \B \\  
6203690227 & 1.74$^{+0.09}_{-0.10}$ &0.872$\pm$0.101 &$<$0.26&- & - & 0.54$\pm$0.03 &52(72)\T \B \\  
6203690228 & 1.71$\pm$0.06 &0.627$\pm$0.016 &$<$0.25&- & - & 0.40$\pm$0.01 &89(78)\T \B \\  
6203690229 & 1.92$^{+0.03}_{-0.04}$ &1.016$\pm$0.030 &$<$0.18&- & - & 0.52$\pm$0.01 &77(88)\T \B \\

\hline
\hline
\end{longtable}

\clearpage

\begin{table}
\begin{center}
\caption{Best fit values derived from simultaneously modelling the \swift-XRT spectra using Model 0. The $N_{\rm H}$ was frozen to 0.9$\times10^{22}$\,cm$^{-2}$. $K_{\rm nthc}$ is the normalisation of the \texttt{nthcomp} component. In the fit, the $kT_{\rm seed}$ parameter was fixed to 0.1\,keV. The flux is unabsorbed and in the 0.5--10\,keV energy interval. 
\label{tab:spec_swift}}
\begin{tabular}{ccccc}
\hline
\hline
Obs. & $\Gamma$ & $K_{\rm nthc}$ & $F_{0510}$ & $\chi^2$ (dof) \\ 
     &  & $(\times100)$ & ($\times$10$^{-11}$\,\unitF) & \\ 
\hline 
1 & 1.83$\pm$0.05 & 6.53$\pm$0.29 & 36$\pm$1 & 41.42(48) \\          
2 & 2.13$\pm$0.05 & 5.01$\pm$0.24 & 21.5$\pm$0.4 & 161.75(145) \\
3 & 1.97$\pm$0.06 & 3.02$\pm$0.18 & 14.7$\pm$0.4 & 83.56(86) \\
4 & 2.07$\pm$0.05 & 3.57$\pm$0.18 & 16.1$\pm$0.3 & 141.22(130) \\
5 & 1.88$\pm$0.06 & 1.76$\pm$0.12 & 9.3$\pm$0.3 & 93.61(79) \\
6 & 1.78$\pm$0.06 & 1.33$\pm$0.09 & 7.8$\pm$0.3 & 68.45(65) \\
7 & 1.72$\pm$0.10 & 0.92$\pm$0.10 & 5.7$\pm$0.3 & 44.18(27) \\
8 & 1.73$\pm$0.11 & 0.77$\pm$0.09 & 4.7$\pm$0.3 & 29.93(30) \\
9 & 1.55$\pm$0.14 & 0.64$\pm$0.10 & 4.8$\pm$0.4 & 4.62(14) \\
10 & 1.45$\pm$0.14 & 0.47$\pm$0.07 & 3.9$\pm$0.3 & 16.08(15) \\ 
11 & 1.24$\pm$0.18 & 0.25$\pm$0.05 & 2.9$\pm$0.4 & 8.06(13) \\
12 & 1.32$\pm$0.09 & 0.34$\pm$0.04 & 3.4$\pm$0.2 & 32.16(29) \\ 
13 & $<$1.11 & 0.17$^{+0.03}_{-0.01}$ & 2.9$\pm$0.3 & 5.28(10) \\
14 & 1.20$\pm$0.11 & 0.21$\pm$0.03 & 2.5$\pm$0.2 & 16.97(18) \\
15 & 1.36$\pm$0.13 & 0.26$\pm$0.04 & 2.5$\pm$0.2 & 12.71(18) \\
16 & 1.35$\pm$0.16 & 0.22$\pm$0.03 & 2.1$\pm$0.2 & 19.44(20) \\
16 & 1.24$\pm$0.14 & 0.16$\pm$0.01 & 1.8$\pm$0.2 & 14.09(21) \\
18 & $<$1.41 & 0.05$\pm$0.01 & 0.6$\pm$0.1 & 5.31(12) \\
19 & $<$1.22 & 0.04$\pm$0.01 & 0.6$^\pm$0.1 & 9.60(10) \\
20 & $<$1.28 & 0.07$\pm$0.02 & 1.1$\pm$0.2 & 13.99(14) \\
21 & 1.46$\pm$0.14 & 0.27$\pm$0.04 & 2.3$\pm$0.2 & 36.73(42) \\
22 & 1.75$\pm$0.13 & 0.83$\pm$0.11 & 5.0$\pm$0.4 & 63.02(64) \\
23 & 1.75$\pm$0.13 & 1.49$\pm$0.19 & 9.0$\pm$0.6 & 50.03(57) \\
24 & 1.69$\pm$0.15 & 1.29$\pm$0.20 & 8.3$\pm$0.7 & 24.58(40) \\
25 & 1.69$\pm$0.16 & 1.57$\pm$0.25 & 10$\pm$1 & 25.80(36) \\
26 & 1.89$\pm$0.18 & 1.81$\pm$0.30 & 9$\pm$1 & 24.37(34) \\
27 & 1.50$\pm$0.17 & 0.82$\pm$0.16 & 6.6$\pm$0.8 & 25.17(25) \\
28 & 1.59$\pm$0.12 & 0.69$\pm$0.09 & 4.9$\pm$0.4 & 63.87(62) \\
29 & 1.54$\pm$0.20 & 0.61$\pm$0.13 & 4.6$\pm$0.6 & 18.23(21) \\      
30 & 1.86$\pm$0.19 & 1.14$\pm$0.21 & 6.1$\pm$0.6 & 40.38(30) \\
31 & 1.91$\pm$1.36 & 2.21$^{+0.70}_{-0.63}$ & 11$\pm$2 & 6.11(8) \\
32 & 1.76$\pm$0.10 & 1.39$\pm$0.15 & 8.3$\pm$0.5 & 77.04(87) \\
33 & 1.82$\pm$0.11 & 1.48$\pm$0.17 & 8.3$\pm$0.5 & 57.25(75) \\
34 & 1.83$\pm$0.12 & 1.07$\pm$0.13 & 5.9$\pm$0.4 & 52.77(62) \\
35 & 1.84$\pm$0.11 & 0.95$\pm$0.10 & 5.3$\pm$0.3 & 97.30(83) \\
36 & 1.94$\pm$0.15 & 1.24$\pm$0.18 & 6.3$\pm$0.5 & 38.69(40) \\
37 & 1.78$\pm$0.16 & 0.71$\pm$0.11 & 4.2$\pm$0.4 & 33.20(39) \\ 
38 & 1.86$\pm$0.10 & 0.89$\pm$0.09 & 4.8$\pm$0.3 & 62.00(85) \\
39 & 1.58$\pm$0.13 & 0.49$\pm$0.06 & 3.5$\pm$0.3 & 34.67(54) \\
40 & 1.77$\pm$0.11 & 0.68$\pm$0.07 & 4.0$\pm$0.2 & 77.90(84) \\
   
\hline
\hline
\end{tabular}
\end{center}
\end{table}

\end{document}